\documentclass[
preprint,
amsmath,amssymb,
aps, physrev,
prd,
longbibliography,superscriptaddress,nofootinbib,floatfix,superscriptaddress,amssymb
]{revtex4}
\usepackage{caption}
\usepackage{graphicx}
\usepackage{subcaption}%
\usepackage{dcolumn}
\usepackage{bm}
\usepackage{float}

\begin{document}
	
	\title{Spiral momentum spectra and spin asymmetry in vacuum pair production driven by phase-modulated elliptically polarized electric fields}
	
	\author{Li-Yi Yang}
	\affiliation{Key Laboratory of Beam Technology of the Ministry of Education, and School of Physics and Astronomy, Beijing Normal University, Beijing 100875, China}
	\author{Hong-Hao Fan}
	\affiliation{School of Mathematics and Physics, Xinjiang Hetian College, Hetian 848000, China}
	\author{Jing-He Huang}
	\affiliation{Key Laboratory of Beam Technology of the Ministry of Education, and School of Physics and Astronomy, Beijing Normal University, Beijing 100875, China}
	\author{Shuo Yang}
	\affiliation{Key Laboratory of Beam Technology of the Ministry of Education, and School of Physics and Astronomy, Beijing Normal University, Beijing 100875, China}
	\author{Bai-Song Xie \footnote{Corresponding Author: bsxie@bnu.edu.cn}}
	\affiliation{Key Laboratory of Beam Technology of the Ministry of Education, and School of Physics and Astronomy, Beijing Normal University, Beijing 100875, China}
	\affiliation{Institute of Radiation Technology, Beijing Academy of Science and Technology, Beijing 100875, China}
	\date{\today}
	
	\begin{abstract}
		We study vacuum electron--positron pair production in phase-modulated elliptically polarized electric fields using the quantum kinetic equations. We analyze how the modulation amplitude, frequency, ellipticity, and modulation form affect the spin-resolved momentum spectra and the spin asymmetry. Cosine-type modulation can deform ring-shaped momentum distributions into spiral patterns. This deformation depends nonmonotonically on the modulation amplitude and frequency and is most pronounced for circular polarization. In the transverse momentum plane, the spin-resolved yields favor the spin channel along the direction of field handedness. Reversing the field handedness would reverse the favored channel while leaving the total yield unchanged. Sine-type modulation preserves ring-shaped and mirror-symmetric spectra approximately, however, it leads to a stronger dependence of the spin asymmetry on the modulation amplitude than cosine-type modulation does. These results show that phase modulation and ellipticity can control the momentum spectra patterns in vacuum pair production. Moreover, phase modulation can also affect the degree of spin asymmetry.
	\end{abstract}

	
	\maketitle
	\section{Introduction}\label{sec:intro}
	
	Electron--positron pair production from vacuum is a typical nonperturbative process in strong-field quantum electrodynamics (QED)~\cite{sauter1931,heisenberg1936,schwinger1951,di2012extremely,fedotov2023advances,xie2017electron}. The rate is exponentially suppressed below the Schwinger critical field $E_{\mathrm{cr}}\approx1.3\times10^{16}\,\mathrm{V/cm}$~\cite{marinov1977electron}. In time-dependent fields, multiphoton transitions allow pair production at subcritical field strengths~\cite{brezin1970pair}. The E-144 experiment at the Stanford Linear Accelerator Center provided an important demonstration of multiphoton strong-field QED~\cite{burke1997}. High-intensity laser facilities have further stimulated studies of pair production in time-dependent fields~\cite{mourou2006,turcu2015strong,kostyukov2023international,ringwald2001pair}. Several theoretical approaches have been developed, including quantum kinetic theory, the Dirac--Heisenberg--Wigner (DHW) formalism, worldline instantons, and other semiclassical methods~\cite{vasak1987quantum,bialynicki1991phase,bickers2014,hebenstreit2010schwinger,dunne2006worldline,dunne2006multidimensional,blinne2016comparison,aleksandrov2024kinetic}. The temporal structure of the field can enhance pair production and produce interference effects in the momentum spectra~\cite{schutzhold2008dynamically,dunne2009catalysis,linder2015pulse,otto2015lifting,dumlu2010stokes,akkermans2012ramsey}.
	
	The momentum distributions of produced particles contain information about the temporal structure, polarization, and interference dynamics of the driving field~\cite{dumlu2011interference,li2017vortices,hu2023spirals,
		olugh2019pair}. Interference between pair-production
	contributions at different times can produce oscillatory
	structures in the momentum spectra~\cite{dumlu2011interference,akkermans2012ramsey}. Spiral and vortex structures have been studied in strong-field ionization and
	photodetachment~\cite{macek2009origin,ngoko2015electron,pengel2017electron,ji2020trajectories,majczak2022carrier}. Related structures appear in vacuum pair production driven by rotating and counter-rotating fields~\cite{hu2024spin}. In bichromatic counter-rotating fields, the relative time delay controls the number of spiral arms~\cite{li2017vortices,hu2023spirals}. Frequency chirping also modifies interference effects and substantially reshapes the momentum spectra in circularly polarized fields~\cite{olugh2019pair}. These studies establish rotating electric fields as a suitable setting for investigating the interplay among field rotation, interference, and spin in pair production, and such fields have been analyzed using semiclassical and Wigner-function approaches~\cite{blinne2014pair,strobel2015semiclassical}.
	Circular polarization is particularly suitable for spin-resolved studies because the rotating field couples particle motion to spin dynamics~\cite{kohlfuerst2018}. In circularly polarized fields, spin-resolved momentum spectra with spiral structures have been reported~\cite{chen2025spin}, providing information on the production dynamics beyond spin-summed distributions~\cite{jiang2026study}. 
	
	Despite these developments, the combined effects of phase modulation and ellipticity on spin-resolved momentum spectra in spatially homogeneous elliptically polarized fields have not been systematically investigated. Phase modulation has been
	studied in spatially inhomogeneous fields~\cite{hu2023pair}, and pair production in fields with both spatial and temporal dependence has also been investigated~\cite{ruf2009pair,hebenstreit2011particle}. A linear chirp changes the carrier frequency monotonically. In contrast, sinusoidal phase modulation produces a periodically varying instantaneous frequency without changing the pulse envelope, thereby reshaping the interference pattern. Ellipticity determines the rotational character of the transverse field and adjusts the balance between transverse rotation and directional anisotropy. Because the two factors act simultaneously on the spectral geometry and on the spin precession, their combined influence cannot be inferred from either factor alone.
	
	In this work, we study vacuum electron--positron pair production in phase-modulated elliptically polarized electric fields using the quantum kinetic equations (QKEs). We calculate spin-summed and spin-resolved momentum distributions for different modulation amplitudes,
	frequencies, ellipticities, and forms of phase modulation. For circular polarization, cosine-type modulation transforms ring-shaped distributions into spiral patterns. The deformation does not increase monotonically with the modulation amplitude or frequency. It peaks at intermediate values within the sampled range and is most clearly resolved in this case. Varying the ellipticity changes the angular deformation and the anisotropy of the momentum distribution. Sine-type modulation preserves ring-shaped, approximately mirror-symmetric distributions, but its spin asymmetry exhibits a much stronger dependence on the modulation amplitude compared with the cosine-type case. For circular polarization with \(\delta=+1\), the spin-down channel is favored throughout the ranges of the modulation amplitude and frequency considered. Reversing the field handedness reverses the favored channel while leaving the total yield unchanged.
	
	This paper is organized as follows. Section~\ref{sec:theory} introduces the
	QKEs and the field model. Section~\ref{sec:results} discusses how the momentum spectra, contour deformation, and spin-resolved yields and spin asymmetry depend on the modulation amplitude, frequency, and ellipticity.
	Section~\ref{sec:phase_modulation_form} compares the momentum spectra and
	spin asymmetry for cosine-type and sine-type phase modulations.
	Section~\ref{sec:summary} summarizes the main results and provides an outlook.

	\section{Theoretical Framework}\label{sec:theory}
	
	We consider vacuum electron--positron pair production in a spatially homogeneous time-dependent elliptically polarized electric field. Within the dipole approximation, the field is
	\begin{equation}
		\bm{E}(t)
		=
		\frac{E_0 e^{-t^2/(2\tau^2)}}{\sqrt{1+\delta^2}}
		\begin{pmatrix}
			\cos\Phi(t) \\
			\delta\sin\Phi(t) \\
			0
		\end{pmatrix},
	\end{equation}
	where $E_0$ is the field-strength parameter, $\tau$ the pulse duration, and $\Phi(t)$ the phase function. Here $\delta=0$ and $|\delta|=1$ correspond to linear and circular polarization, respectively, and the sign of \(\delta\) sets the rotation sense. For \(\delta>0\), the field rotates counterclockwise when viewed along the \(+z\) axis; \(\delta<0\) corresponds to the opposite rotation sense. The factor \(1/\sqrt{1+\delta^2}\) keeps the cycle-averaged intensity fixed as \(\delta\) varies. For \(0\le|\delta|\le1\), the peak field amplitude is \(E_0/\sqrt{1+\delta^2}\).
	
	We consider two forms of phase modulation,
	\begin{equation}
		\Phi_{\cos}(t)
		=
		b\cos(\Omega t)+\omega_0t,
		\qquad
		\Phi_{\sin}(t)
		=
		b\sin(\Omega t)+\omega_0t,
	\end{equation}
	where \(\omega_0\) is the carrier frequency, \(b\) the modulation amplitude, and \(\Omega\) the modulation frequency. The cosine-type modulation serves as the reference configuration, while the sine-type modulation is introduced to examine the role of the temporal phase structure. The corresponding instantaneous angular frequencies are
	\begin{equation}
		\omega_{\mathrm{inst}}^{\cos}(t)
		=
		\frac{\mathrm{d}\Phi_{\cos}}{\mathrm{d}t}
		=
		\omega_0-b\Omega\sin(\Omega t),
		\qquad
		\omega_{\mathrm{inst}}^{\sin}(t)
		=
		\frac{\mathrm{d}\Phi_{\sin}}{\mathrm{d}t}
		=
		\omega_0+b\Omega\cos(\Omega t).
	\end{equation}
	
	To formulate the QKEs, we introduce the vector potential in the temporal gauge, $A^0=0$, for which
	\begin{equation}
		\bm{E}(t)=-\dot{\bm{A}}(t).
	\end{equation}
	The integration constant in $\bm{A}(t)$ is chosen according to the temporal symmetry of each field configuration. The kinetic momentum is
	\begin{equation}
		\bm{p}(t)=\bm q-e\bm A(t),
	\end{equation}
	where \(\bm q\) is the canonical momentum and \(e\) is the elementary charge. We use natural units with $\hbar=c=m=e=1$.
	
	The pair production process in a spatially homogeneous electric field can be described by the QKEs~\cite{hebenstreit2010schwinger,aleksandrov2024kinetic}, which are exactly equivalent to the DHW formalism in this limit~\cite{vasak1987quantum,bialynicki1991phase}. Following the standard reduction, the Wigner function is independent of the spatial coordinate, and only 10 equal-time components are nonvanishing. We organize these components as a scalar distribution \(f\), a spin vector \(\bm{F}\), and two coherence vectors \(\bm{u}\) and \(\bm{v}\). The QKEs then read
	\begin{align}
		\dot{f}&=-2\bm{\mu}_2\cdot\bm{u},\\
		\dot{\bm{F}}&=2\bm{\mu}_1\times\bm{F}
		-2\bm{\mu}_2\times\bm{v},\\
		\dot{\bm{u}}&=2\bm{\mu}_1\times\bm{u}
		+\bm{\mu}_2(2f-1)+2\varepsilon_{\bm p}\bm{v},\\
		\dot{\bm{v}}&=2\bm{\mu}_1\times\bm{v}
		-2\bm{\mu}_2\times\bm{F}-2\varepsilon_{\bm p}\bm{u},
	\end{align}
	where
	\begin{equation}
		\bm{\mu}_1=
		\frac{\bm{p}\times\bm{E}}{2\varepsilon_{\bm p}(\varepsilon_{\bm p}+1)},
		\qquad
		\bm{\mu}_2=
		\frac{\bm{p}(\bm{E}\cdot\bm{p})
			-\varepsilon_{\bm p}(\varepsilon_{\bm p}+1)\bm{E}}
		{2\varepsilon_{\bm p}^2(\varepsilon_{\bm p}+1)},
	\end{equation}
	with \(\varepsilon_{\bm p}=\sqrt{\bm{p}^2+1}\). Here \(f\) is the spin-averaged particle distribution and \(\bm{F}\) carries the spin-dependent information. The initial conditions correspond to the vacuum, \(f=\bm{F}=\bm{u}=\bm{v}=0\).
	
	The spin-resolved distributions are obtained from the covariant spin projection of the Wigner function~\cite{strobel2015semiclassical,blinne2016comparison,jiang2026study}. We choose the \(z\) axis perpendicular to the polarization plane as the spin-quantization axis, for which the two spin channels can have different number densities~\cite{jiang2026study}. The spin-resolved momentum distributions at a sufficiently late time \(t_{\mathrm{out}}\) are
	\begin{equation}
		f_{\uparrow}(\bm q)
		=
		f(\bm q,t_{\mathrm{out}})
		-
		F_z(\bm q,t_{\mathrm{out}}),
		\qquad
		f_{\downarrow}(\bm q)
		=
		f(\bm q,t_{\mathrm{out}})
		+
		F_z(\bm q,t_{\mathrm{out}}),
	\end{equation}
	where \(F_z>0\) corresponds to a larger spin-down population. Here \(f_{\uparrow,\downarrow}\) denote the spin-resolved distributions of the produced electrons. The positron spin projection is opposite along the same quantization axis. Their sum defines the spin-summed distribution,
	\begin{equation}
		f_{\mathrm{total}}(\bm q)
		=
		f_{\uparrow}(\bm q)
		+
		f_{\downarrow}(\bm q)
		=
		2f(\bm q,t_{\mathrm{out}}).
	\end{equation}
	The momentum distributions below are evaluated at \(q_z=0\) for the transverse spectra.

	The resulting system of ordinary differential equations is solved
	numerically for the 10 Wigner components together with the three
	components of the vector potential. We use a variable-order
	Adams--Bashforth--Moulton solver with adaptive step size, setting the
	relative and absolute tolerances to \(10^{-14}\) and \(10^{-16}\),
	respectively. The time integration runs from \(t=-10\tau\) to
	\(t=10\tau\), where the field has effectively vanished. The transverse momentum plane is sampled on a uniform \(150\times150\) grid over \(q_x,q_y\in[-1.5m,1.5m]\), corresponding to a grid spacing of \(\Delta q=0.02m\). Test runs with a finer grid and a larger cutoff produced consistent results.

	\section{Spin-resolved momentum spectra and spin asymmetry}\label{sec:results}
	We first study vacuum electron--positron pair production for
	circular polarization, setting $\delta=1$ and using the cosine-type phase modulation. The field-strength parameter, carrier frequency, and pulse duration are $E_0=0.1\sqrt{2}E_{\rm cr}$, $\omega_0=0.55\,m$, and $\tau=30/m$, respectively. These parameters correspond to a subcritical field in the multiphoton regime and are of the same order as those used in previous studies~\cite{hu2023spirals,hu2024spin}. All frequencies and momenta below are given in units of $m$.

	\subsection{Modulation amplitude $b$}\label{subsec:modulation_amplitude} 
\begin{figure}[htbp]
		\centering
		\captionsetup{justification=raggedright, singlelinecheck=false}
		\setlength{\tabcolsep}{0pt}
		\makebox[\textwidth][c]{%
			\begin{minipage}{0.35\textwidth}
				\centering
				\includegraphics[width=\textwidth]{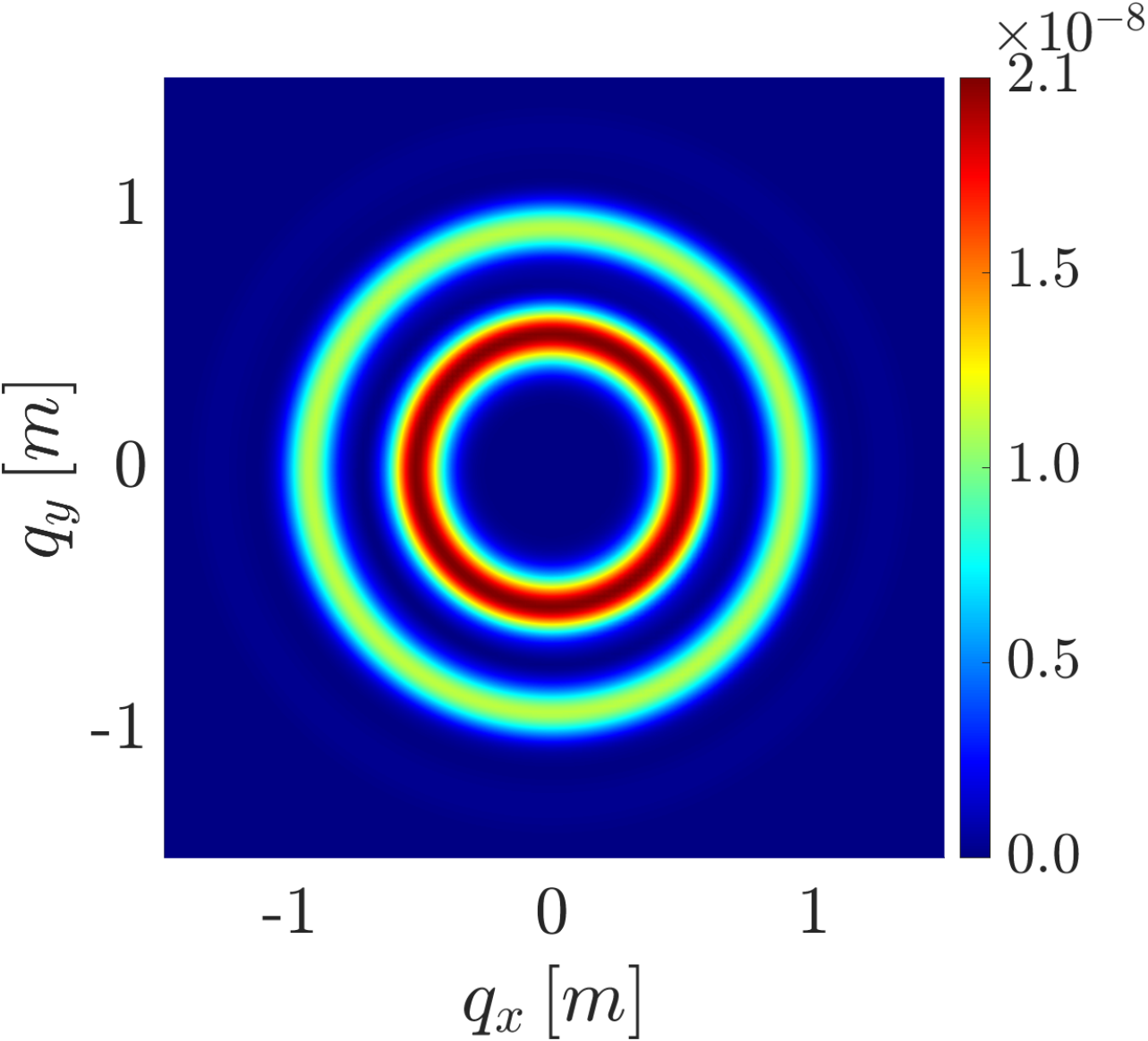}
				\vspace{-0.5em}
				{\small (a) $b=1$}
			\end{minipage}
			\hspace{0.04\textwidth}
			\begin{minipage}{0.35\textwidth}
				\centering
				\includegraphics[width=\textwidth]{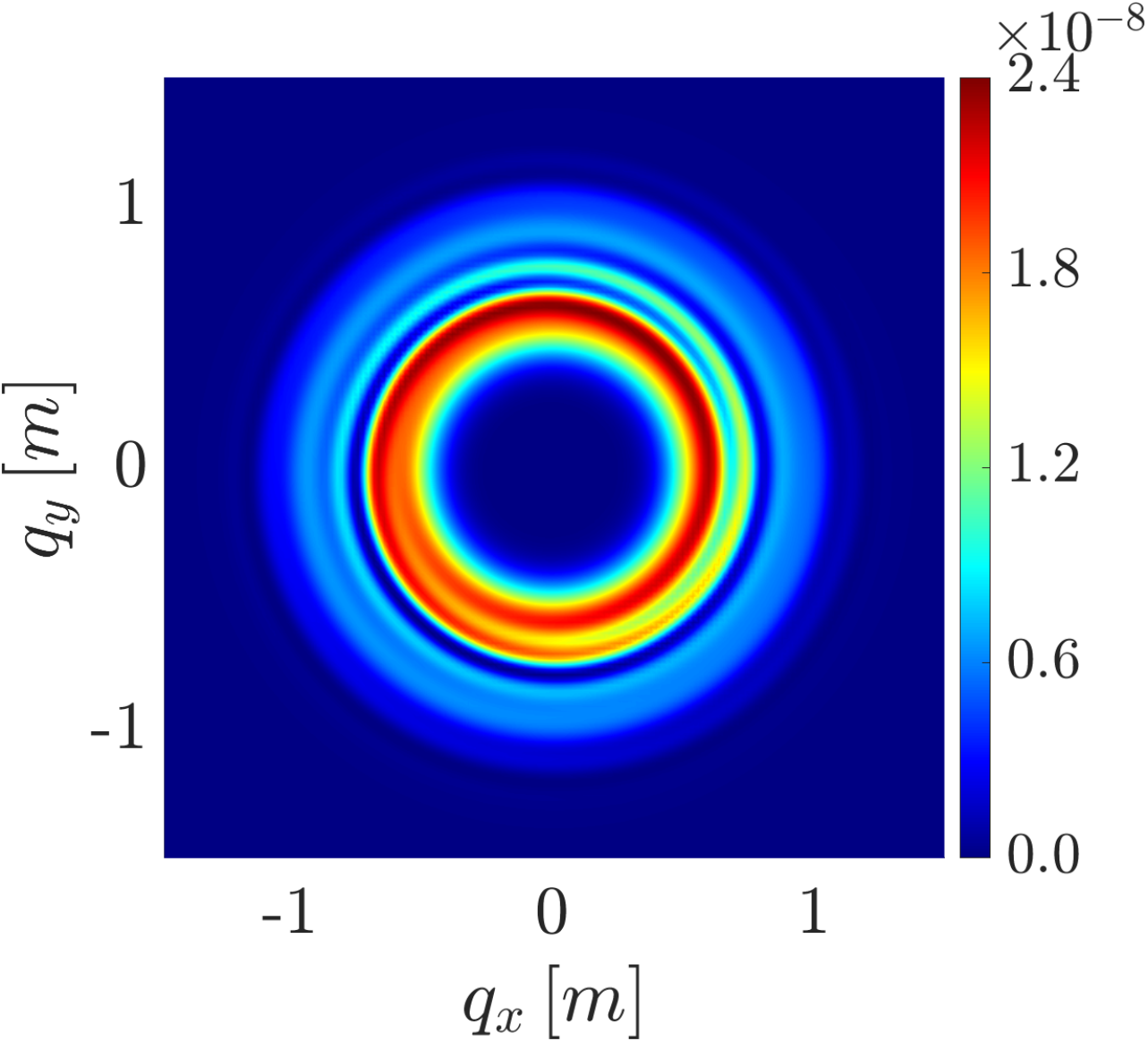}
				\vspace{-0.5em}
				{\small (b) $b=2$}
			\end{minipage}%
		}
		
		\vspace{0.8em}
		
		\makebox[\textwidth][c]{%
			\begin{minipage}{0.35\textwidth}
				\centering
				\includegraphics[width=\textwidth]{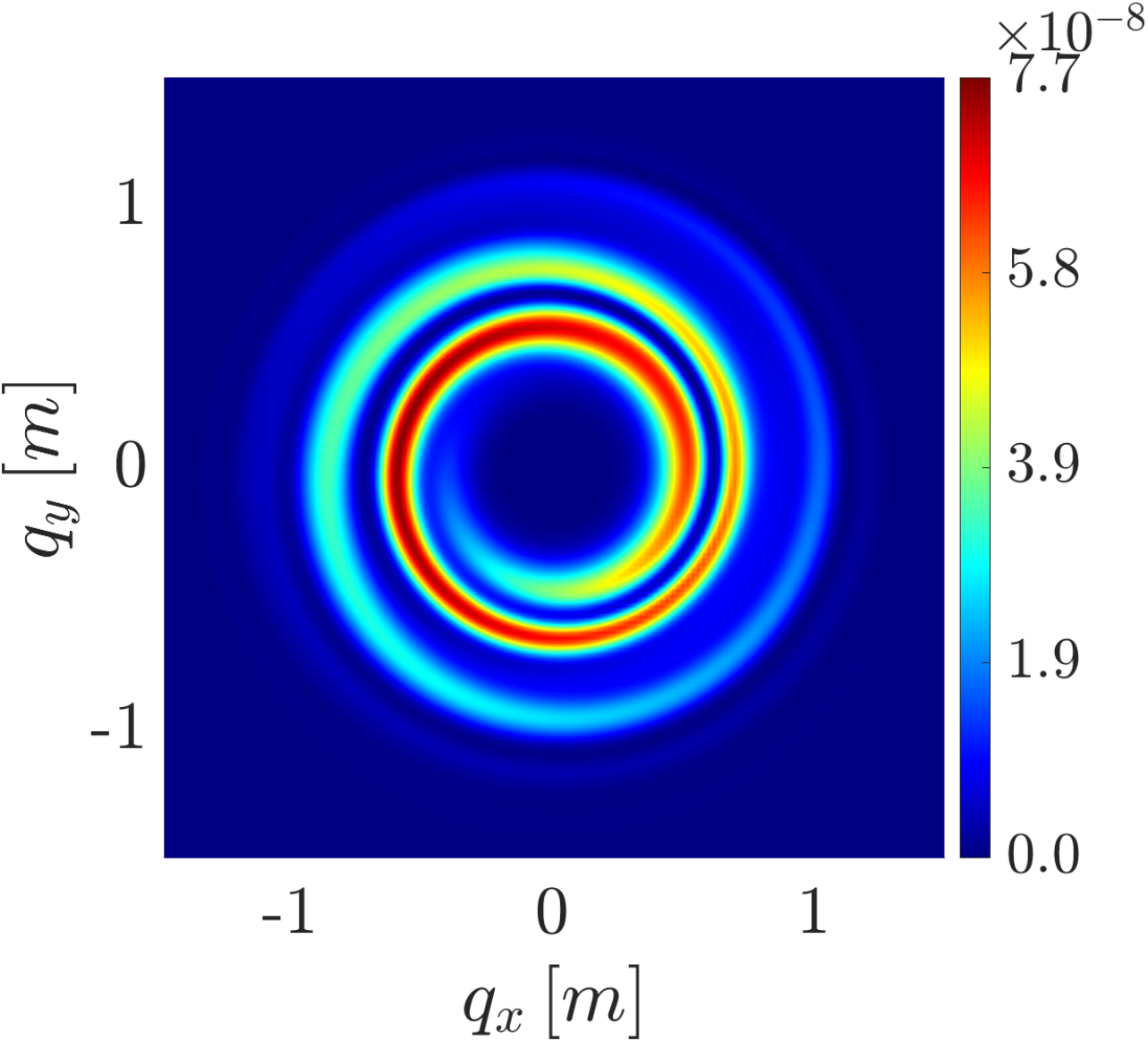}
				\vspace{-0.5em}
				{\small (c) $b=6$}
			\end{minipage}
			\hspace{0.04\textwidth}
			\begin{minipage}{0.35\textwidth}
				\centering
				\includegraphics[width=\textwidth]{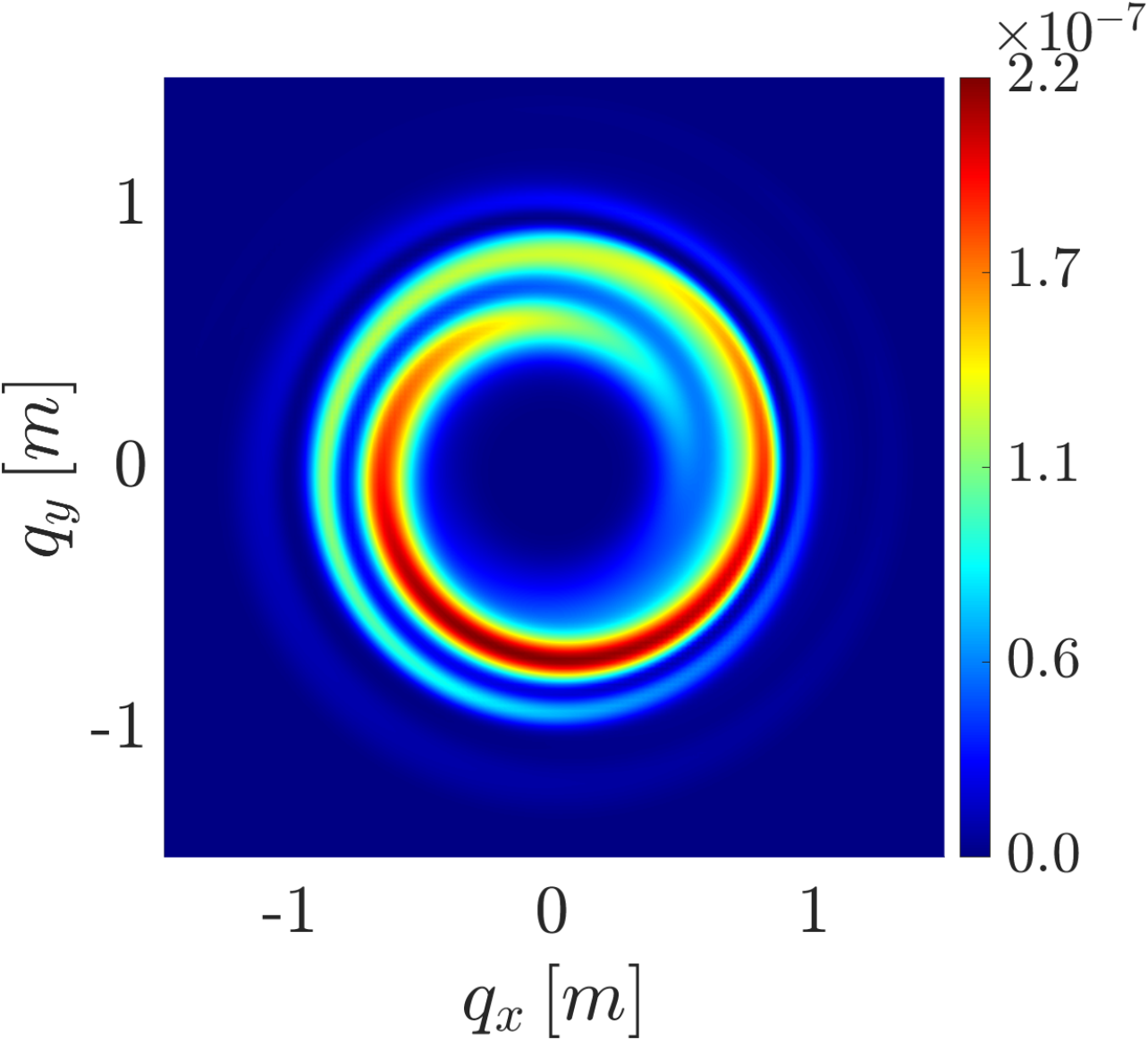}
				\vspace{-0.5em}
				{\small (d) $b=9$}
			\end{minipage}%
		}

		\caption{Transverse momentum distributions of spin-up particles for different modulation amplitudes in a circularly polarized field with cosine-type phase modulation. The modulation amplitudes are $b=1$, $2$, $6$, and $9$. The other parameters are $E_0=0.1\sqrt{2}E_{\rm cr}$, $\omega_0=0.55\,m$, $\tau=30/m$, $\Omega=0.04$, and $q_z=0$.}
		\label{fig:1}
	\end{figure}

	We fix \(\Omega=0.04\) and vary \(b=1,2,6,9\). The spin-up momentum distributions are shown in Fig.~\ref{fig:1}.
	For $b=1$, the distribution is nearly ring-shaped. As $b$ increases to 2 and 6, it develops a pronounced spiral pattern. At \(b=9\), the distribution remains strongly deformed, but its contour becomes more complex.
	
	To characterize this deformation, we extract isodensity contours, which directly reflect the angular distortion of the momentum distribution. The contours are sampled over 150 isodensity levels equally spaced between $1\%$ and $100\%$ of the maximum density. The results from different levels are then combined through the weighted average. In polar coordinates, the radial distance $r$ and polar angle $\theta$ are defined as
	\begin{equation}
		r=\sqrt{q_x^2+q_y^2}, \qquad \theta=\arg(q_x+iq_y),
	\end{equation}
	where $q_x$ and $q_y$ are the transverse momentum components. Using these coordinates, each extracted contour segment is fitted by
	\begin{equation}
		r(\theta)=a_i\theta+c_i,
	\end{equation}
	where \(a_i\) approximates the local slope \(\mathrm{d}r/\mathrm{d}\theta\)
	within the \(i\)-th segment. The sign of $a_i$ gives the local direction of the radial change with the polar angle.
	
	Segments with fewer than eight distinct polar-angle samples are
	discarded. For the remaining segments, a linear fit is retained only if its angular span $\Delta\theta_i$ is nonzero and its coefficient of determination satisfies $R_i^2\geq0.30$. These thresholds exclude segments too short or too poorly resolved to define a reliable local slope, while retaining enough segments for a meaningful weighted average. After the eight-sample selection, the number of remaining segments is denoted by \(N_{\rm seg}\), and the number of segments retained after the linear-fit criteria is denoted by \(N_{\rm fit}\). The weighted mean slope is \(\bar{a} = \sum_i w_i a_i / \sum_i w_i\), with weights \(w_i = \Delta\theta_i R_i^2\), where the sum runs over retained segments. This weighting gives more weight to segments with larger angular spans and better linear fits. The weighted mean coefficient of determination is \(\overline{R^2} = \sum_i w_i R_i^2 / \sum_i w_i\). While \(\bar{a}\) characterizes the local radial change of the contour with respect to the polar angle, \(\overline{R^2}\) measures the average quality
	of the local linear fits. For convenience, we also list $2\pi|\bar{a}|$, which estimates the total radial change of the contour over a full azimuthal turn. This procedure does not assume that the entire
	contour forms a single spiral. The local linear fitting results for the extracted contours are summarized in Table~\ref{tab:1}. The last column of Table~\ref{tab:1} gives the sign of $\bar{a}$: $-$, $+$, and $0$ correspond to $\mathrm{d}r/\mathrm{d}\theta<0$, $\mathrm{d}r/\mathrm{d}\theta>0$, and a negligibly small $|\bar{a}|$, respectively.
	
	\begin{table}[htbp]
		\centering
		\caption{Local linear contour-fitting results for different modulation amplitudes.}
		\label{tab:1}
		\footnotesize
		\setlength{\tabcolsep}{8pt}
		\begin{tabular}{ccccccc}
			\hline\hline
			$b$ & $N_{\rm seg}$ & $N_{\rm fit}$ & $\bar{a}$ & $\overline{R^2}$ & $2\pi|\bar{a}|$ & Sign \\
			\hline
			1 & 531 & 136 & $4.73\times10^{-5}$ & 0.4559 & $2.97\times10^{-4}$ & \(0\) \\
			2 & 415 & 226 & $-1.20\times10^{-2}$ & 0.6046 & $7.52\times10^{-2}$ & \(-\) \\
			6 & 252 & 213 & $-3.37\times10^{-2}$ & 0.7928 & $2.12\times10^{-1}$ & \(-\) \\
			9 & 233 & 202 & $-3.28\times10^{-2}$ & 0.6424 & $2.06\times10^{-1}$ & \(-\) \\
			\hline\hline
		\end{tabular}
	\end{table}
	
	The weighted mean slope is close to zero for $b=1$ and becomes clearly negative at $b=2$. Both $|\bar{a}|$ and $2\pi|\bar{a}|$ are largest at $b=6$, where $\overline{R^2}$ is also highest. At $b=9$, the deformation remains strong, but the lower $\overline{R^2}$ is consistent with a more complex contour geometry. The deformation is therefore most pronounced at $b=6$ in the range considered and does not increase monotonically with $b$. At fixed \(\Omega\), increasing \(b\) increases \(|b\Omega|\), the amplitude of the instantaneous-frequency variation, so particles produced at different times acquire different final momenta. At larger $b$, interference effects between contributions from different production times generate contour segments with distinct local orientations, which results in more complex structures and poorer linear fits.
	
\begin{figure}[htbp]
		\centering
		\captionsetup{justification=raggedright, singlelinecheck=false}
		\setlength{\tabcolsep}{0pt}
		\makebox[\textwidth][c]{%
			\begin{minipage}{0.35\textwidth}
				\centering
				\includegraphics[width=\textwidth]{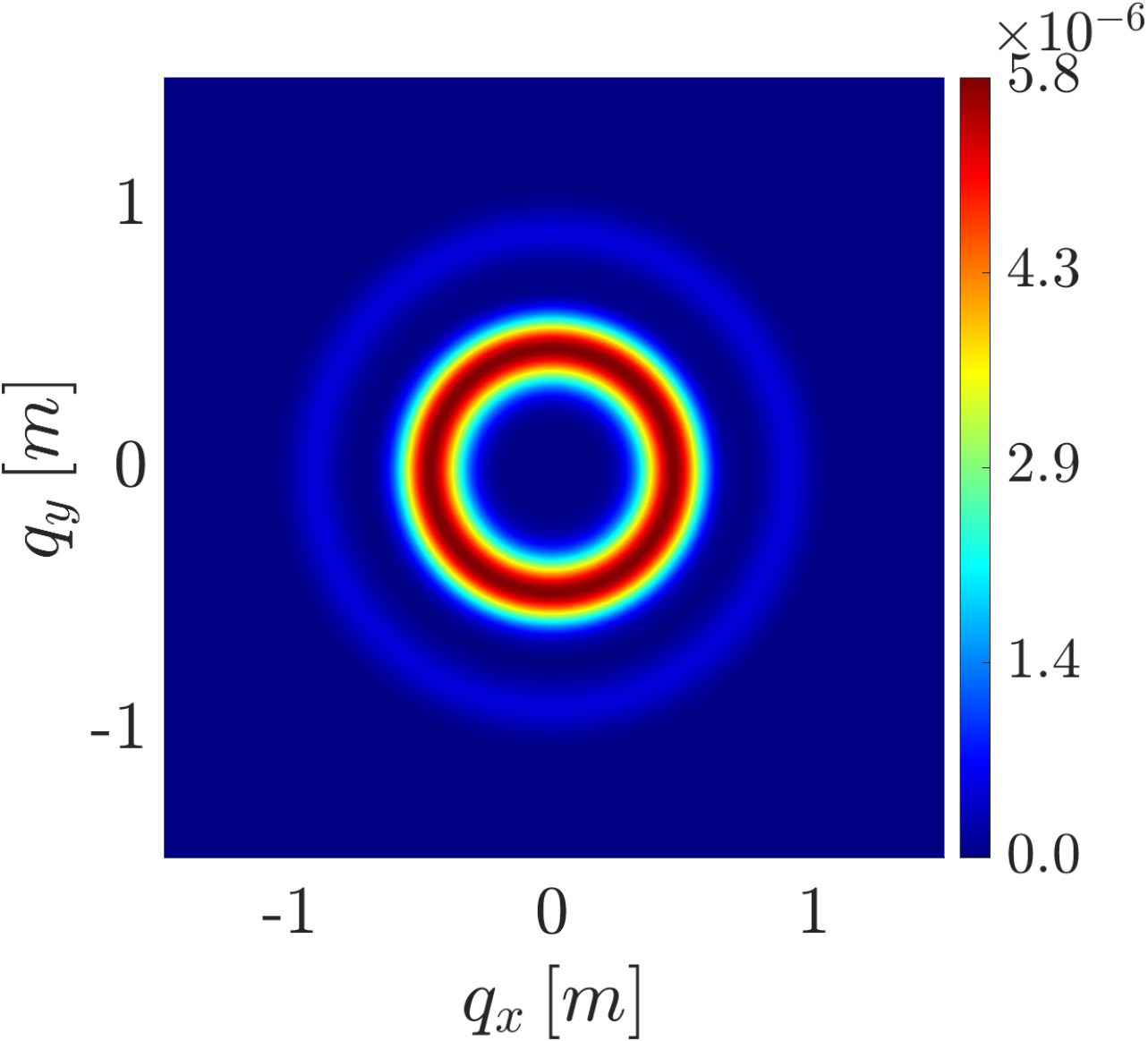}
				\vspace{-0.5em}
				{\small (a) $b=1$}
			\end{minipage}
			\hspace{0.04\textwidth}
			\begin{minipage}{0.35\textwidth}
				\centering
				\includegraphics[width=\textwidth]{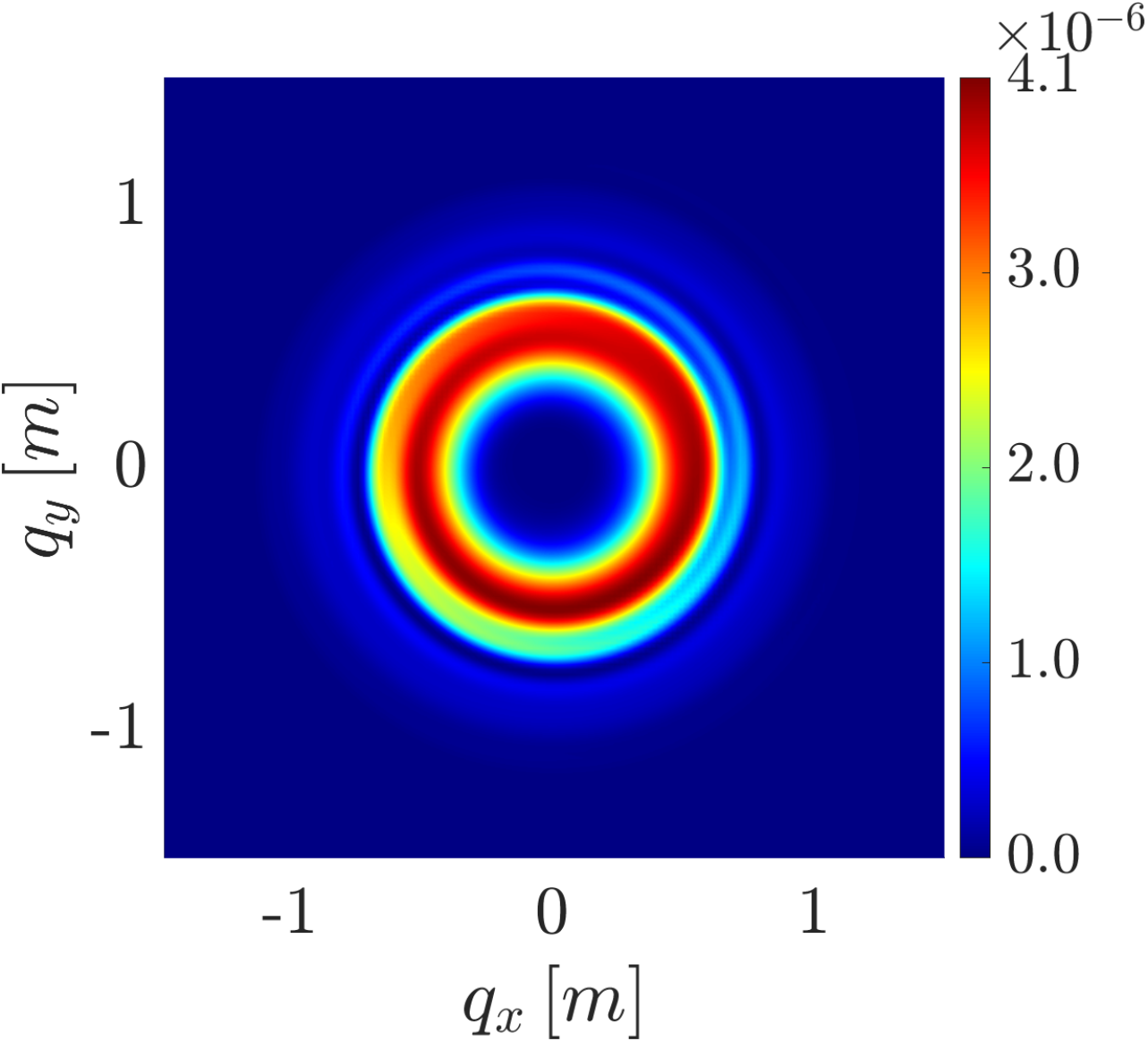}
				\vspace{-0.5em}
				{\small (b) $b=2$}
			\end{minipage}%
		}
		
		\vspace{0.8em}
		
		\makebox[\textwidth][c]{%
			\begin{minipage}{0.35\textwidth}
				\centering
				\includegraphics[width=\textwidth]{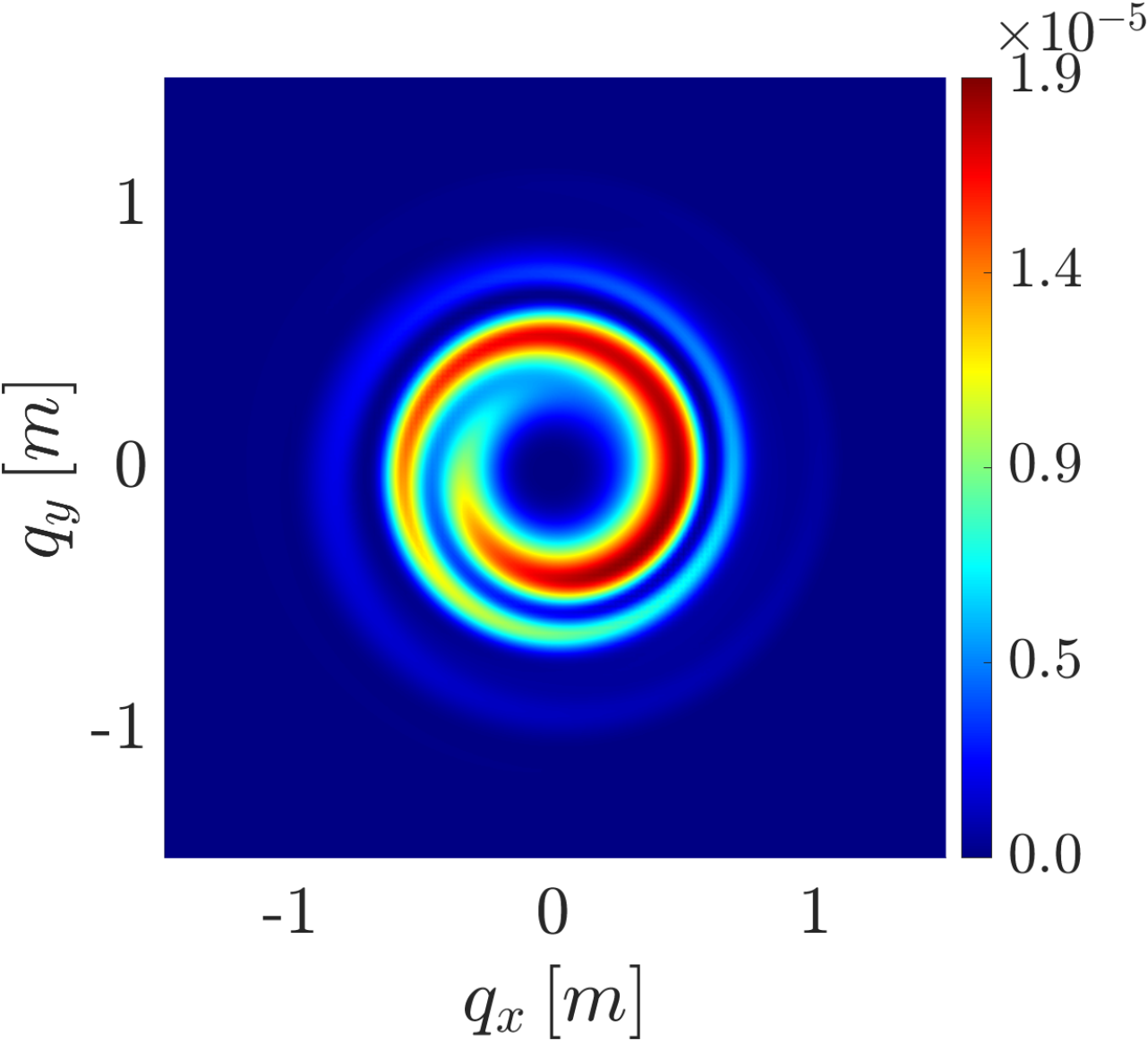}
				\vspace{-0.5em}
				{\small (c) $b=6$}
			\end{minipage}
			\hspace{0.04\textwidth}
			\begin{minipage}{0.35\textwidth}
				\centering
				\includegraphics[width=\textwidth]{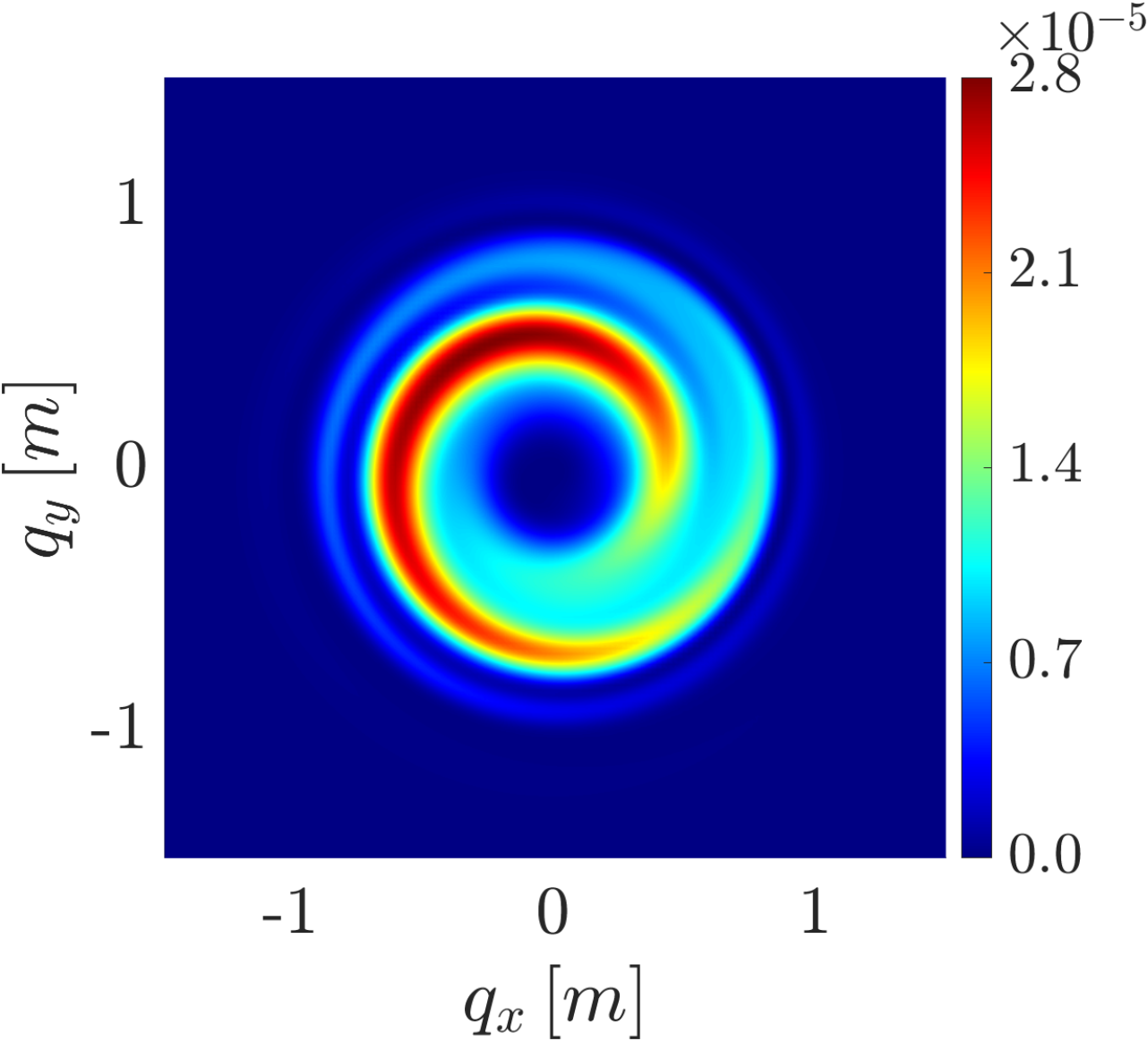}
				\vspace{-0.5em}
				{\small (d) $b=9$}
			\end{minipage}%
		}
		\caption{Transverse momentum distributions of spin-down particles for different modulation amplitudes in a circularly polarized field with cosine-type phase modulation. The parameters are the same as in Fig.~\ref{fig:1}.}
		\label{fig:spin_down}
	\end{figure}

	The spin-down distributions for the same field parameters are shown in Fig.~\ref{fig:spin_down}. Both spin channels exhibit qualitatively similar deformations, indicating that the spiral structure is a property of the pair-production dynamics rather than of a specific spin channel. The spiral pattern is somewhat more pronounced in the spin-up channel, so we use the spin-up distributions for the contour analysis in the following subsections.

	\subsection{Modulation frequency \(\Omega\)}
\begin{figure}[htbp]
		\centering
		\captionsetup{justification=raggedright, singlelinecheck=false}
		\setlength{\tabcolsep}{0pt}
		\makebox[\textwidth][c]{%
			\begin{minipage}{0.35\textwidth}
				\centering
				\includegraphics[width=\textwidth]{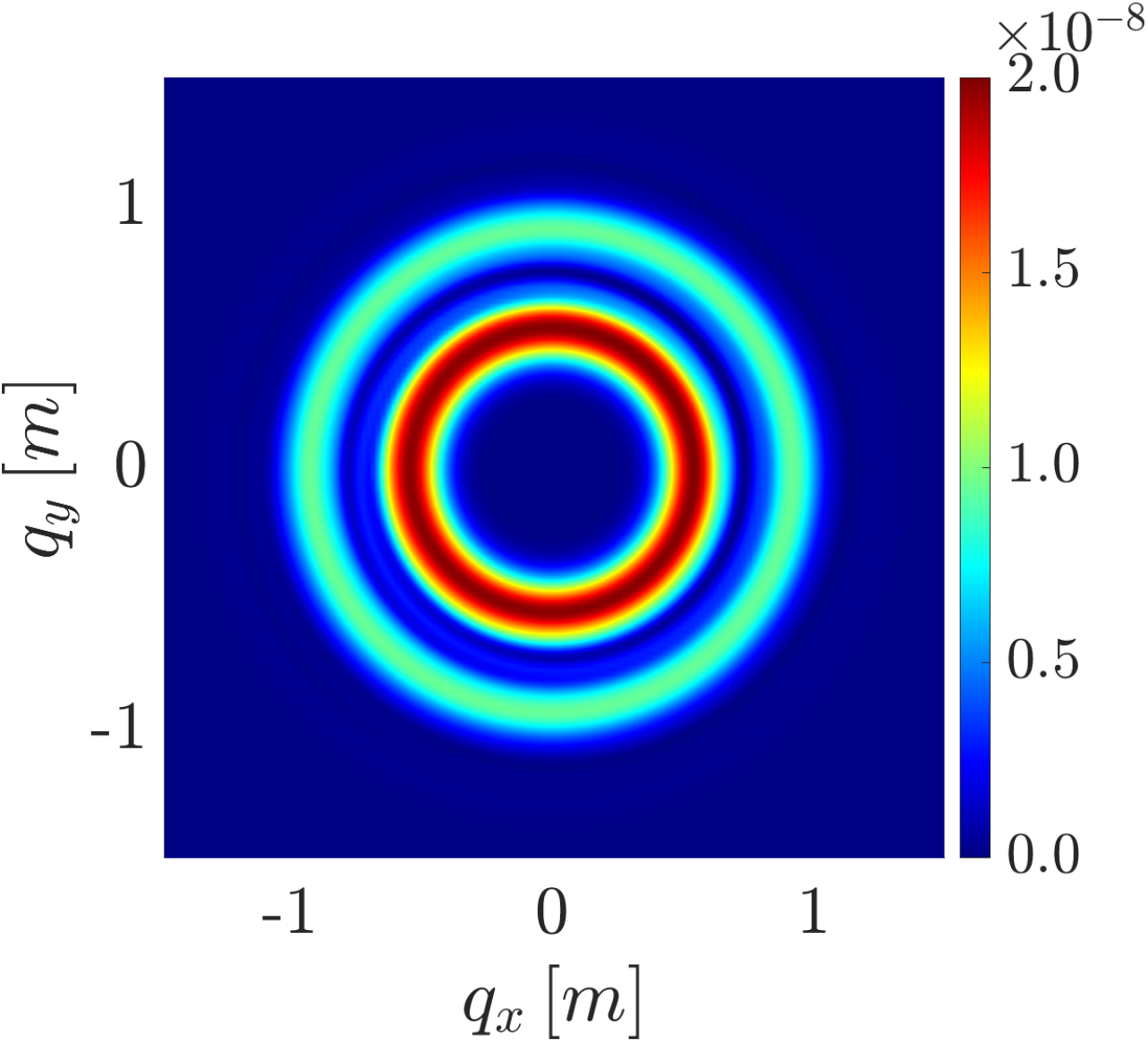}
				\vspace{-0.5em}
				{\small (a) $\Omega=0.02$}
			\end{minipage}
			\hspace{0.04\textwidth}
			\begin{minipage}{0.35\textwidth}
				\centering
				\includegraphics[width=\textwidth]{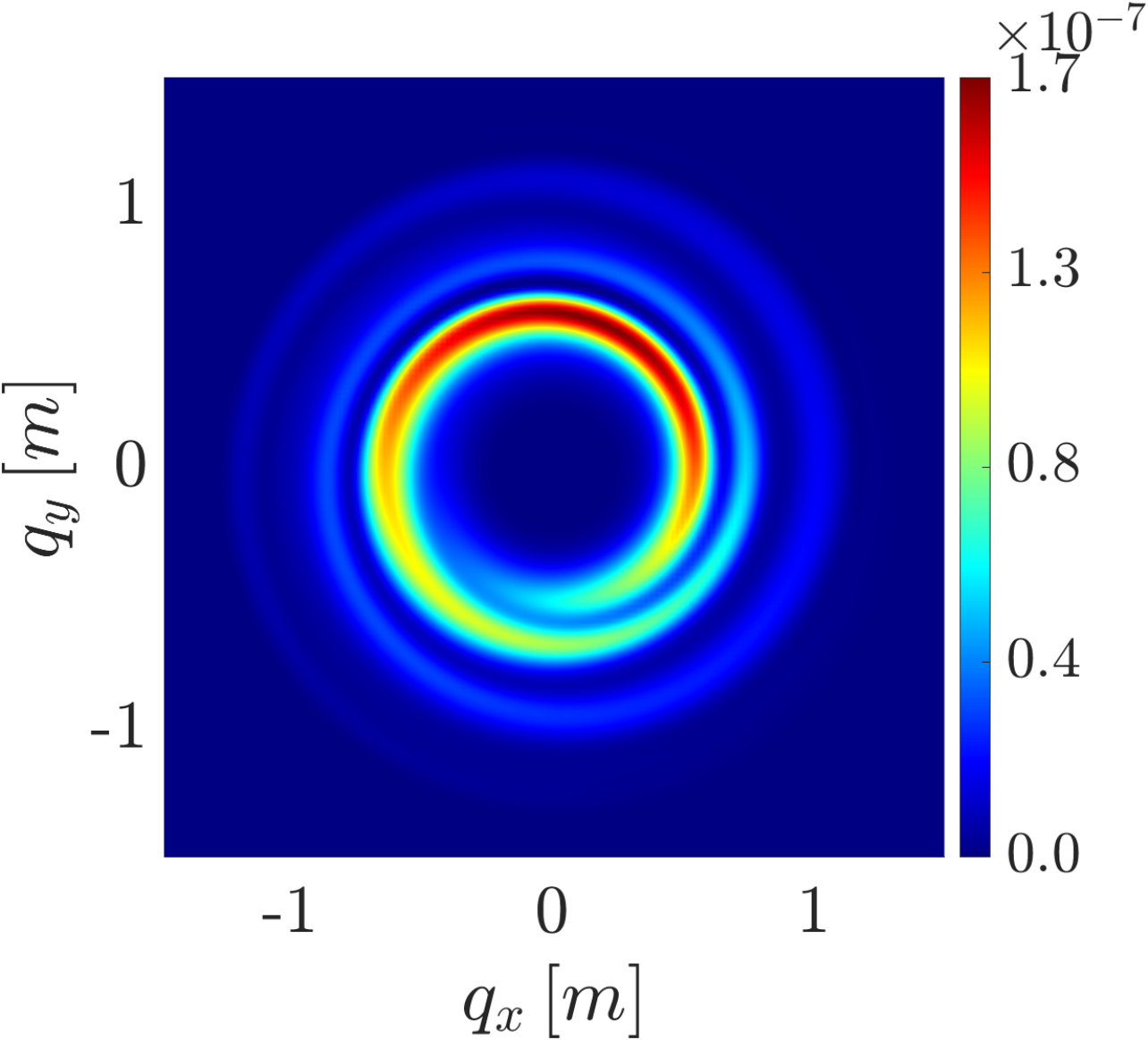}
				\vspace{-0.5em}
				{\small (b) $\Omega=0.05$}
			\end{minipage}%
		}
		
		\vspace{0.8em}
		
		\makebox[\textwidth][c]{%
			\begin{minipage}{0.35\textwidth}
				\centering
				\includegraphics[width=\textwidth]{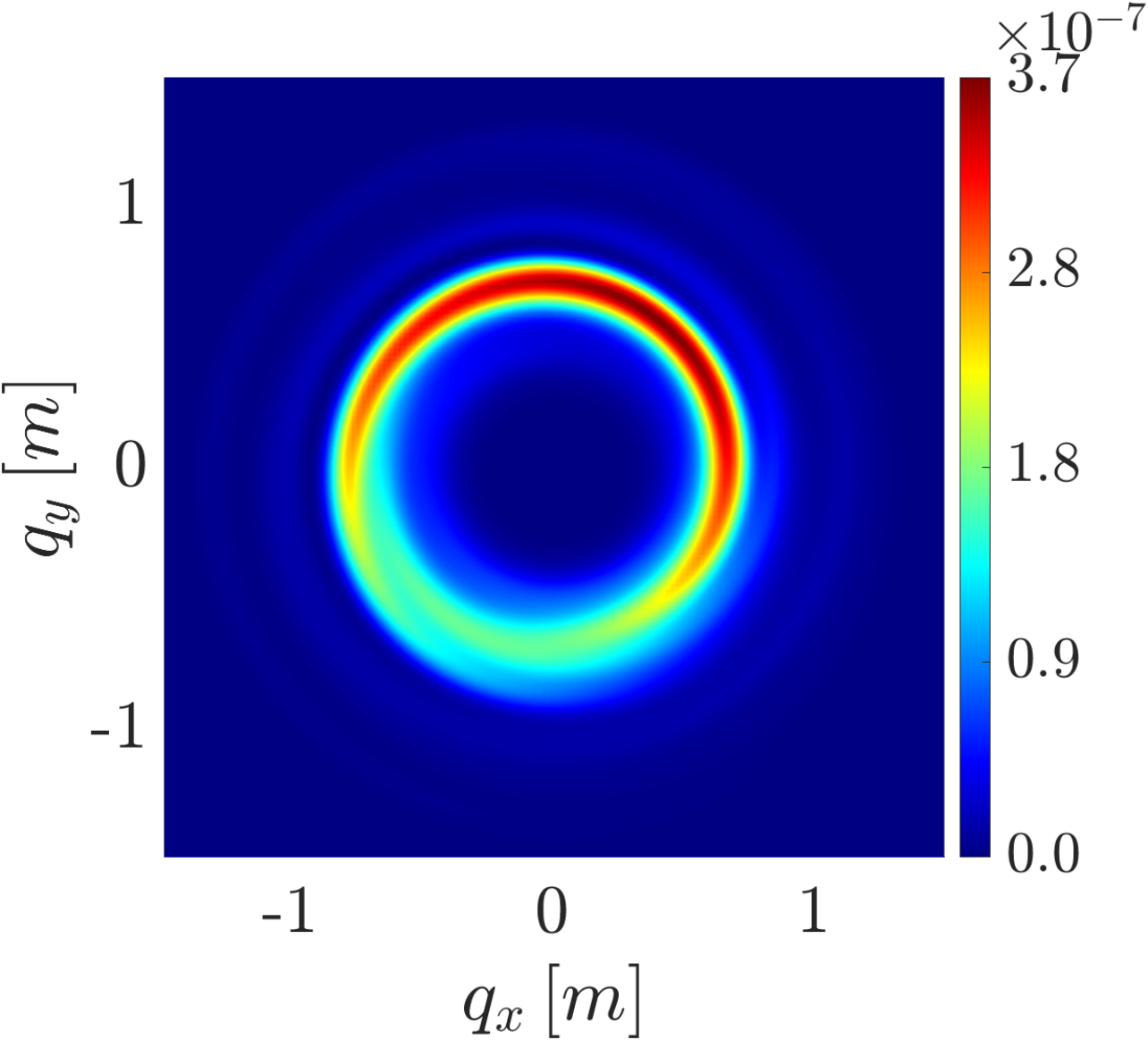}
				\vspace{-0.5em}
				{\small (c) $\Omega=0.06$}
			\end{minipage}
			\hspace{0.04\textwidth}
			\begin{minipage}{0.35\textwidth}
				\centering
				\includegraphics[width=\textwidth]{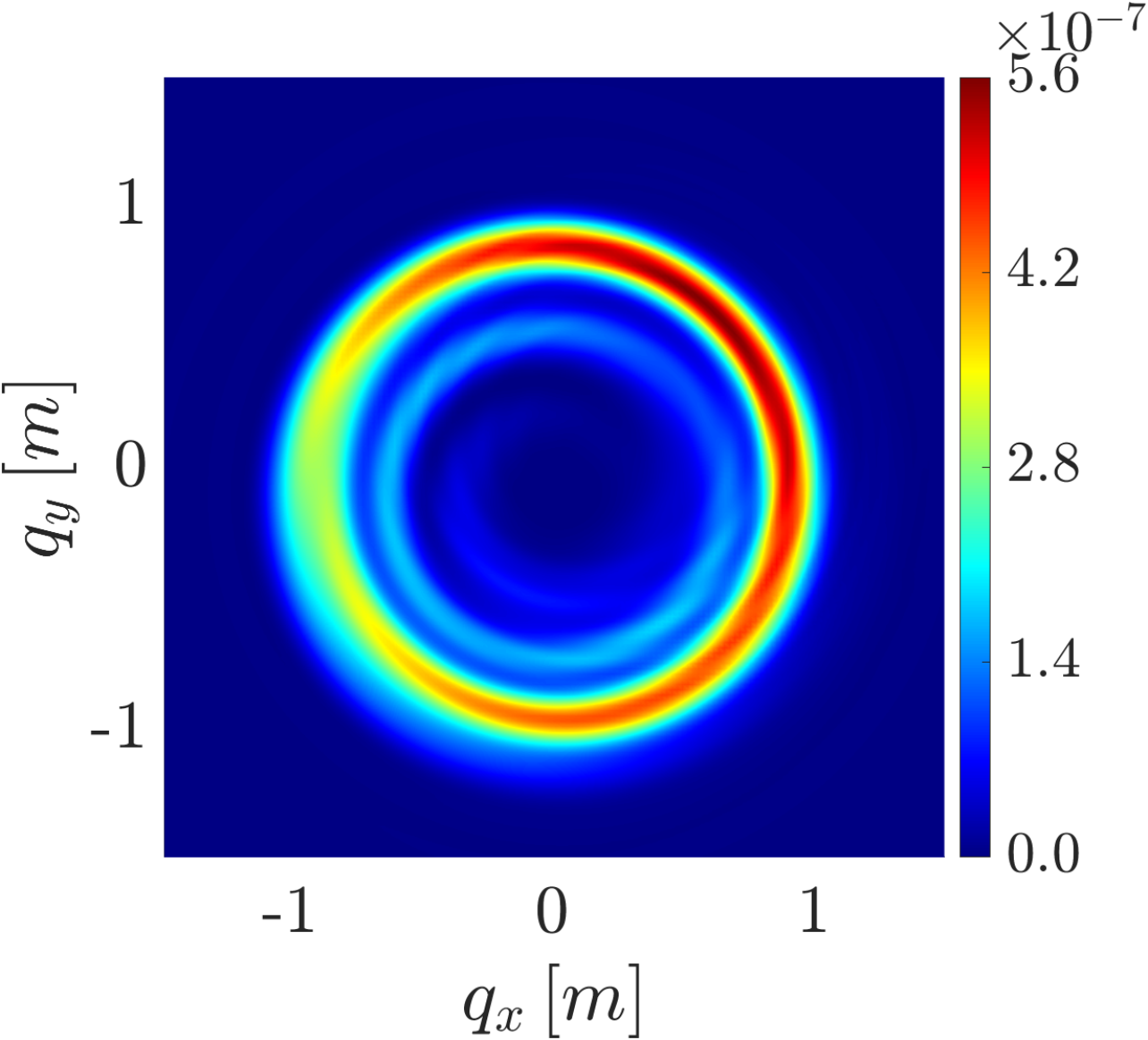}
				\vspace{-0.5em}
				{\small (d) $\Omega=0.08$}
			\end{minipage}%
		}
		
		\caption{Transverse momentum distributions of spin-up particles for different modulation frequencies in a circularly polarized field with cosine-type phase modulation. The modulation frequencies are $\Omega=0.02, 0.05, 0.06, 0.08$, with $b=5$ and $q_z=0$. The other parameters are the same as in Fig.~\ref{fig:1}.}
		\label{fig:3}
	\end{figure}

	We fix \(b=5\) and vary \(\Omega=0.02,0.05,0.06,0.08\). The other parameters are the same as in Fig.~\ref{fig:1}. The spin-up momentum distributions are shown in Fig.~\ref{fig:3}.
	At $\Omega=0.02$, the distribution is close to a ring with only weak deformation. As $\Omega$ increases from 0.02 to 0.05, the ring becomes more distorted and a spiral structure develops. At \(\Omega=0.06\), the deformation weakens. At \(\Omega=0.08\), the net orientation of the contour slopes reverses.
	
	We use the same local contour-fitting procedure as in
	Sec.~\ref{subsec:modulation_amplitude} for the spin-up distributions. The fitting results are listed in Table~\ref{tab:2}.
	\begin{table}[htbp]
		\centering
		\caption{Local linear contour-fitting results for different modulation frequencies.}
		\label{tab:2}
		\footnotesize
		\setlength{\tabcolsep}{8pt}
		\begin{tabular}{ccccccc}
			\hline\hline
			$\Omega$ & $N_{\rm seg}$ & $N_{\rm fit}$ & $\bar{a}$ & $\overline{R^2}$ & $2\pi|\bar{a}|$ & Sign \\
			\hline
			0.02 & 476 & 173 & $-1.71\times10^{-3}$ & 0.5002 & $1.08\times10^{-2}$ & \(0\) \\
			0.05 & 213 & 178 & $-2.51\times10^{-2}$ & 0.5338 & $1.58\times10^{-1}$ & \(-\) \\
			0.06 & 221 & 124 & $-1.74\times10^{-2}$ & 0.4267 & $1.09\times10^{-1}$ & \(-\) \\
			0.08 & 307 & 191 & $8.72\times10^{-3}$ & 0.5350 & $5.48\times10^{-2}$ & \(+\) \\
			\hline\hline
		\end{tabular}
	\end{table}
	The contour deformation varies nonmonotonically with $\Omega$. At $\Omega=0.02$, the weighted mean slope is small, consistent with the nearly ring-shaped distribution. The deformation is strongest at $\Omega=0.05$ and weaker at $\Omega=0.06$, where the lower \(\overline{R^2}\) indicates that the contours are less well described by local linear fits. At $\Omega=0.08$, the sign of \(\bar{a}\) changes, indicating a change in the net direction inferred from the local fits. Varying \(\Omega\) at fixed \(b\) changes the amplitude \(|b\Omega|\) of the instantaneous-frequency variation. Increasing \(|b\Omega|\) makes the instantaneous frequency vary more strongly during the pulse, so particles produced at different times accumulate different phases. This changes the angular distribution of the final momenta and initially enhances the deformation. At sufficiently large \(|b\Omega|\), however, the rapid phase variation can modify the interference pattern and even reverse the net direction of the local contour slopes. 
	
	When \(b\) is varied with \(\Omega=0.04\), the most pronounced spiral appears at \(b=6\), giving \(|b\Omega|=0.24\). When \(\Omega\) is varied with \(b=5\), the strongest deformation occurs at \(\Omega=0.05\), giving \(|b\Omega|=0.25\). The similar values of \(|b\Omega|\) at the peak deformation suggest that this product is a key parameter determining the maximum deformation. Beyond this common dependence, \(\Omega\) also changes the modulation period \(2\pi/\Omega\). Since the pulse duration is fixed, different \(\Omega\) values correspond to different fractions of a modulation cycle being sampled within the pulse. This difference in the sampled phase range can reverse the net direction of the local contour slopes at \(\Omega=0.08\), whereas no such reversal appears over the range of \(b\) considered here. Thus, \(\Omega\) provides an additional control parameter for the orientation of the spiral deformation.

	\subsection{Ellipticity \(\delta\)}
	\label{subsec:ellipticity}
	
\begin{figure}[htbp]
		\centering
		\captionsetup{justification=raggedright, singlelinecheck=false}
		\setlength{\tabcolsep}{0pt}
		\begin{minipage}{0.32\textwidth}
			\centering
			\includegraphics[width=\textwidth]{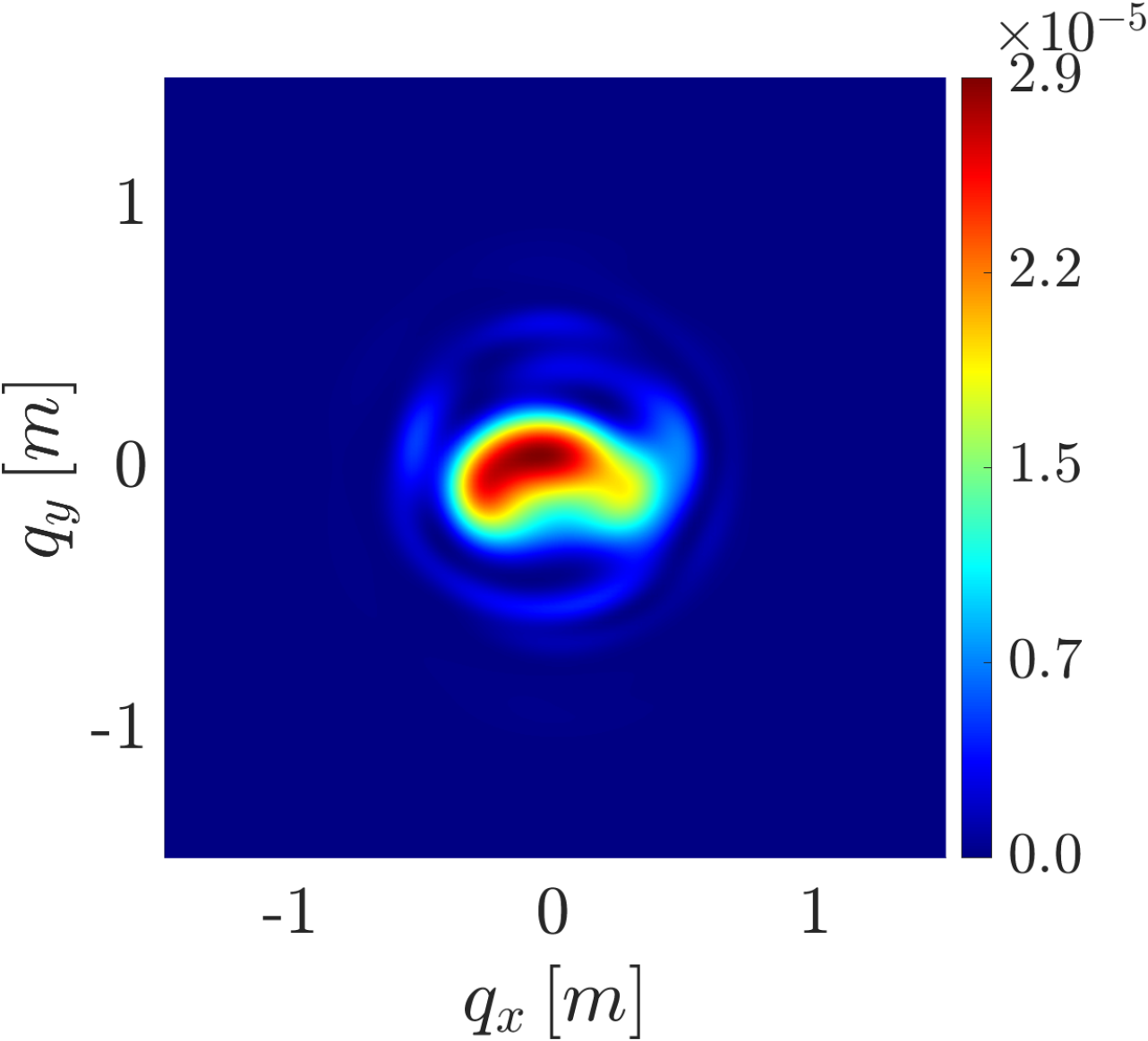}
			\vspace{-0.5em}
			{\small (a) $\delta=0.2$}
		\end{minipage}
		\hfill
		\begin{minipage}{0.32\textwidth}
			\centering
			\includegraphics[width=\textwidth]{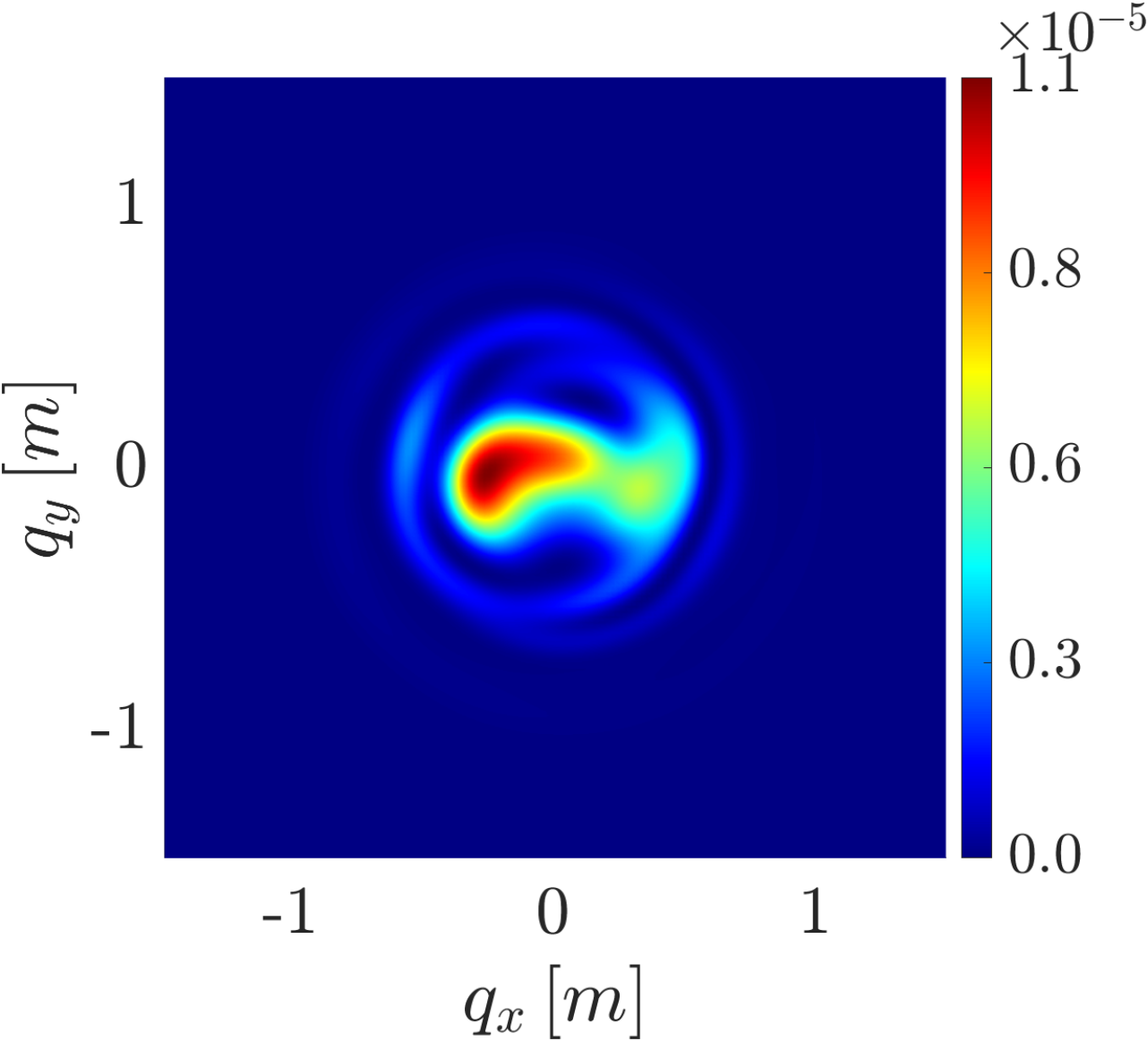}
			\vspace{-0.5em}
			{\small (b) $\delta=0.4$}
		\end{minipage}
		\hfill
		\begin{minipage}{0.32\textwidth}
			\centering
			\includegraphics[width=\textwidth]{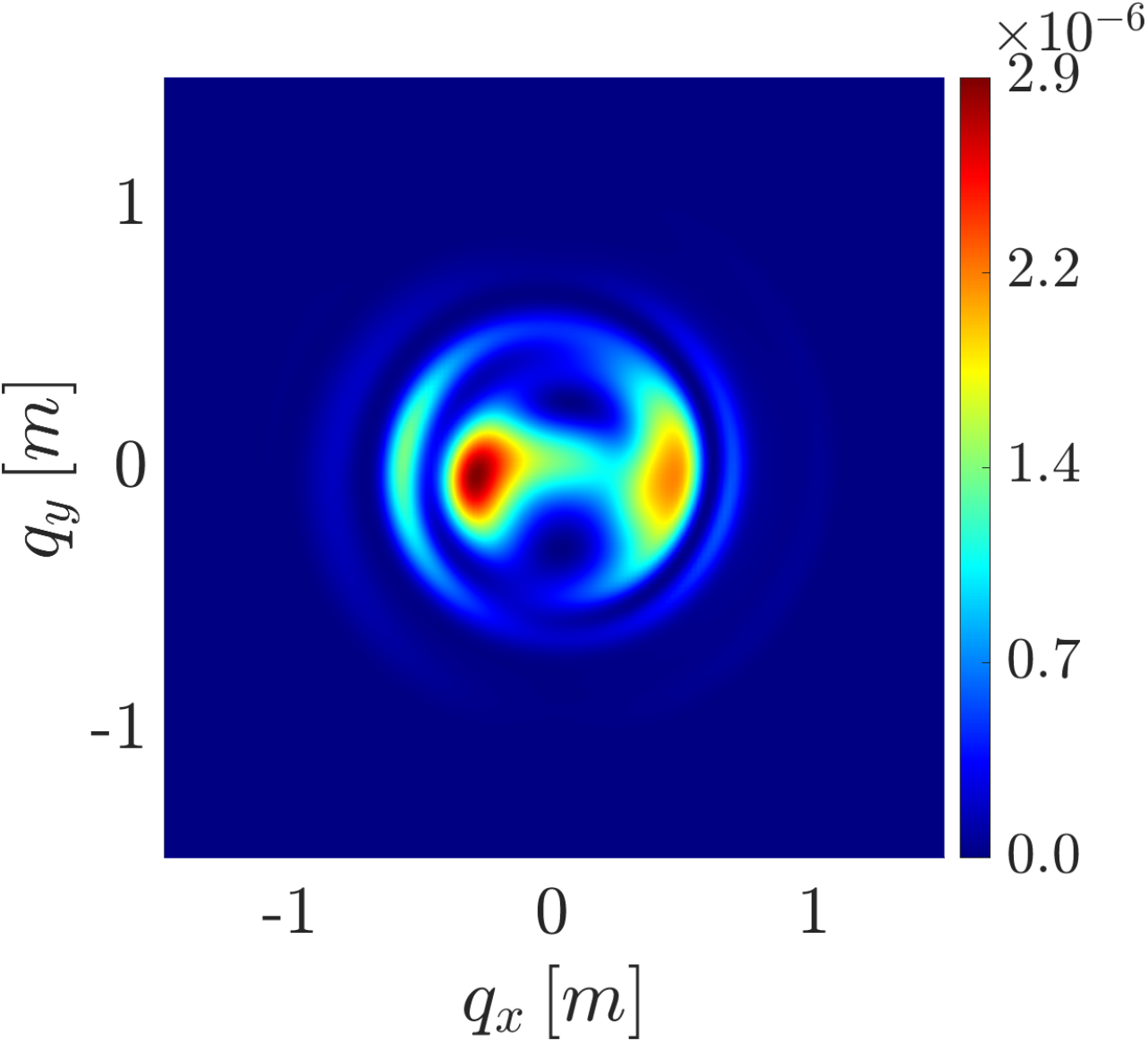}
			\vspace{-0.5em}
			{\small (c) $\delta=0.6$}
		\end{minipage}
		
		\vspace{1em}
		
		\makebox[\textwidth][c]{%
			\begin{minipage}{0.32\textwidth}
				\centering
				\includegraphics[width=\textwidth]{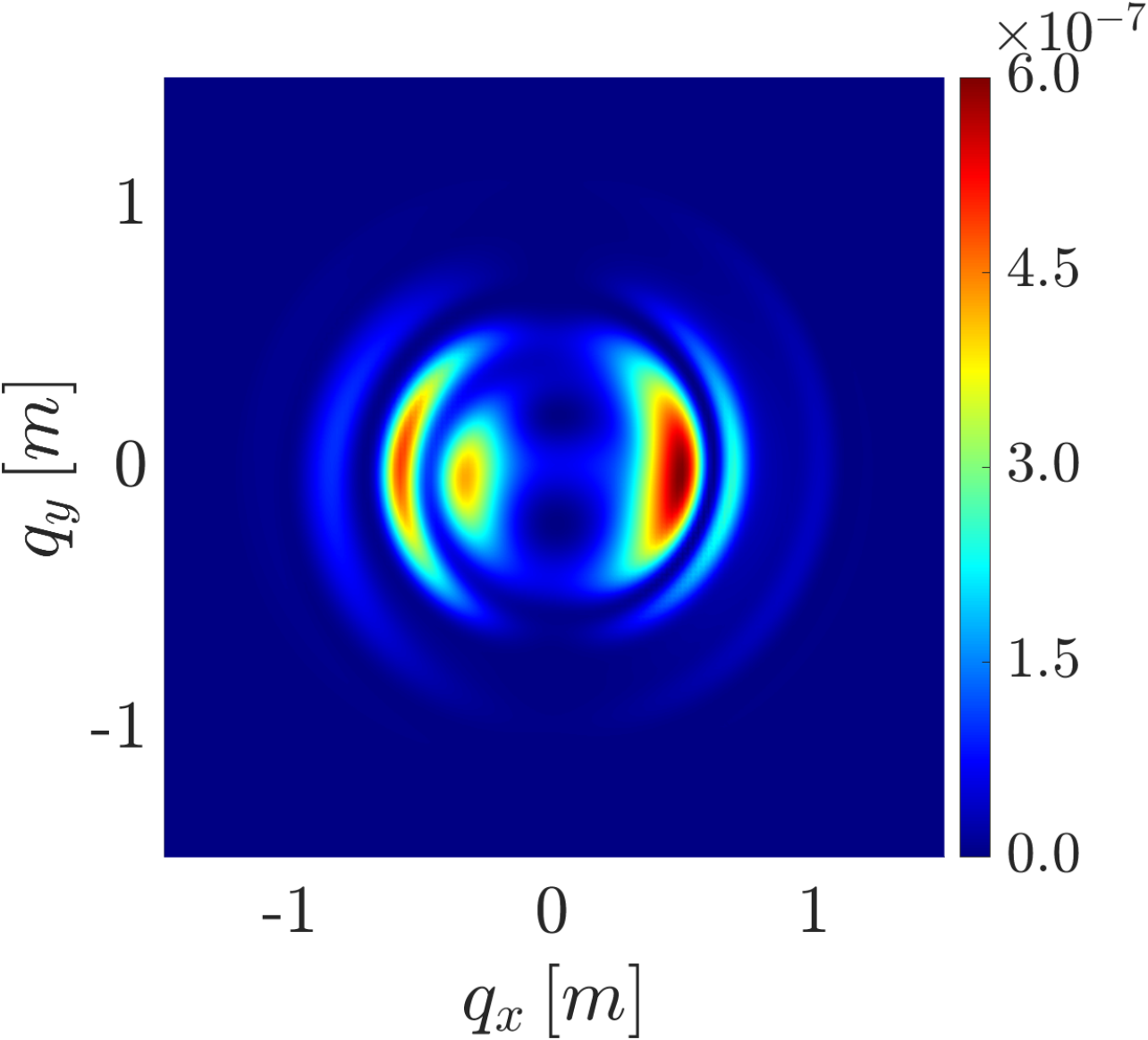}
				\vspace{-0.5em}
				{\small (d) $\delta=0.8$}
			\end{minipage}
			\hspace{0.04\textwidth}
			\begin{minipage}{0.32\textwidth}
				\centering
				\includegraphics[width=\textwidth]{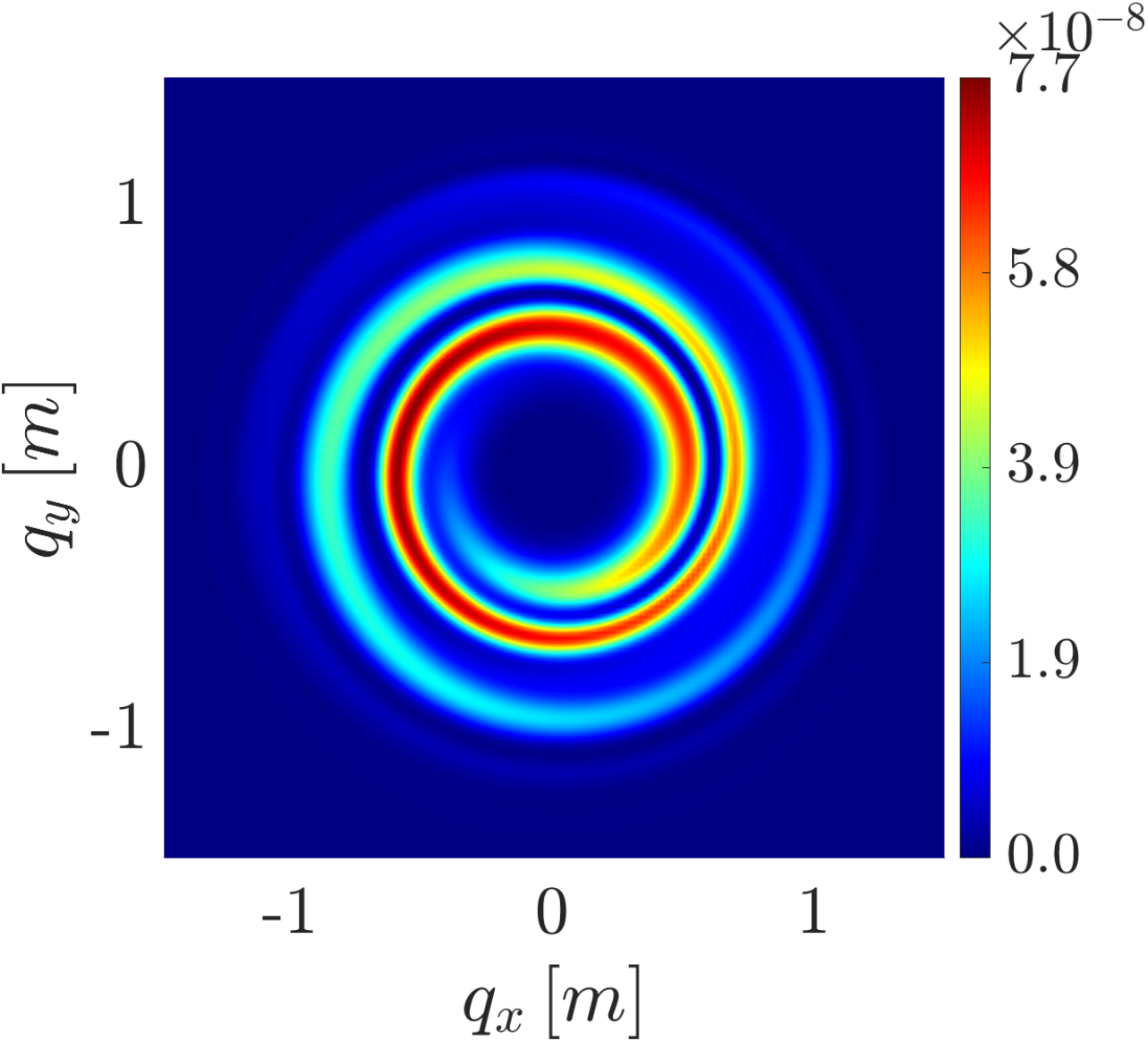}
				\vspace{-0.5em}
				{\small (e) $\delta=1.0$}
			\end{minipage}%
		}
		
		\caption{Transverse momentum distributions of spin-up particles for different ellipticities in a cosine-type phase-modulated field. The ellipticities are $\delta=0.2$, $0.4$, $0.6$, $0.8$, and $1.0$. The modulation parameters are $b=6$ and $\Omega=0.04$, with $q_z=0$. The other parameters are $E_0=0.1\sqrt{2}E_{\rm cr}$, $\omega_0=0.55\,m$, and $\tau=30/m$.}
		\label{fig:ellipticity_spectra} 
	\end{figure}

	We fix \(b=6\) and \(\Omega=0.04\), for which the spiral deformation was
	most pronounced in Sec.~\ref{subsec:modulation_amplitude}, and vary
	\(\delta=0.2,0.4,0.6,0.8,1.0\). The remaining parameters are the same as in Fig.~\ref{fig:1}. The spin-up momentum distributions are shown in Fig.~\ref{fig:ellipticity_spectra}. 
	
	For small \(\delta\), the distribution is strongly anisotropic and
	elongated along the \(q_x\) direction, consistent with the nearly
	linear polarization of the field~\cite{olugh2020asymmetric}. As \(\delta\) increases, the overall distribution becomes less elongated and more ring-like, while the contour develops a spiral deformation.
	
	To quantify the contour deformation, we use the local contour-fitting procedure described above. The fitting results are listed in Table~\ref{tab:ellipticity}. For elliptical fields, \(\bar{a}\) is not a robust global descriptor because the contours are locally anisotropic, and \(N_{\rm fit}\) is much smaller. The circularly polarized case has the largest \(N_{\rm fit}\) and the best mean fit quality, so it gives the most reliable contour description. In particular, only six segments are retained at \(\delta=0.8\), so the corresponding \(\bar{a}\) and \(\overline{R^2}\) values are statistically limited and should not be overinterpreted.

	\begin{table}[htbp]
		\centering
		\caption{Local linear contour-fitting results for different ellipticities.}
		\label{tab:ellipticity}
		\footnotesize
		\setlength{\tabcolsep}{8pt}
		\begin{tabular}{ccccccc}
			\hline\hline
			$\delta$ & $N_{\rm seg}$ & $N_{\rm fit}$ & $\bar{a}$ &
			$\overline{R^2}$ & $2\pi|\bar{a}|$ & Sign \\
			\hline
			0.2 & 205 & 27  & $1.356\times10^{-2}$ & 0.4381 &
			$8.52\times10^{-2}$ & \(+\) \\
			0.4 & 230 & 34  & $-6.620\times10^{-3}$ & 0.4636 &
			$4.16\times10^{-2}$ & \(-\) \\
			0.6 & 329 & 18  & $-2.626\times10^{-2}$ & 0.4087 &
			$1.65\times10^{-1}$ & \(-\) \\
			0.8 & 456 & 6   & $1.690\times10^{-2}$ & 0.5958 &
			$1.06\times10^{-1}$ & \(+\) \\
			1.0 & 252 & 213 & $-3.369\times10^{-2}$ & 0.7928 &
			$2.12\times10^{-1}$ & \(-\) \\
			\hline\hline
		\end{tabular}
	\end{table}
	
	As noted above, the normalization keeps the cycle-averaged intensity fixed while reducing the peak amplitude. The amplitude ratio \(E_y/E_x\) is determined by \(\delta\). As \(\delta\) increases, the \(y\)-component of the transverse motion becomes stronger, and the overall motion becomes more rotational. At \(\delta=1.0\), the two transverse components have equal amplitudes, so the field has no preferred transverse axis. This makes the transverse motion more isotropic, so the modulation-induced phase differences can directly modify the azimuthal distribution, making the spiral pattern most pronounced among the sampled ellipticities. These results indicate that the spiral structure is governed jointly by the phase modulation and the ellipticity. The phase modulation controls the temporal phase evolution and the instantaneous frequency, while the ellipticity controls the balance between transverse rotation and directional anisotropy. Ellipticity thus provides an additional control parameter for the geometry of the transverse momentum spectra.

	\subsection{Spin-resolved yields and spin asymmetry}
	\label{subsec:spin_resolved_yields}
	
	We analyze the spin-resolved yields and spin asymmetry in the \(q_z=0\) transverse momentum plane. The spin-up and spin-down yields are $N_{\uparrow,\downarrow}=\int dq_x\,dq_y\,f_{\uparrow,\downarrow}(q_x,q_y,q_z=0)$. The total yield in this plane is \(N_{\mathrm{total}} = N_{\uparrow} + N_{\downarrow}\). We characterize the spin-resolved yields by the spin asymmetry and the spin-up fraction,
	\begin{equation}
		\mathcal{K}_{\mathrm{total}}     
		=
		\frac{N_{\uparrow}-N_{\downarrow}}
		{N_{\uparrow}+N_{\downarrow}},
		\qquad
		P_{\uparrow}
		=
		\frac{N_{\uparrow}}
		{N_{\uparrow}+N_{\downarrow}}
		\times 100\%.
	\end{equation}
	With this convention, $\mathcal{K}_{\mathrm{total}}>0$ corresponds to $N_{\uparrow} > N_{\downarrow}$, and $\mathcal{K}_{\mathrm{total}}<0$ corresponds to $N_{\downarrow} > N_{\uparrow}$.

\begin{figure}[htbp]
		\centering
		\captionsetup{justification=raggedright, singlelinecheck=false}
		\setlength{\tabcolsep}{0pt}
		\begin{subfigure}[b]{0.44\linewidth}
			\centering
			\includegraphics[width=\linewidth]{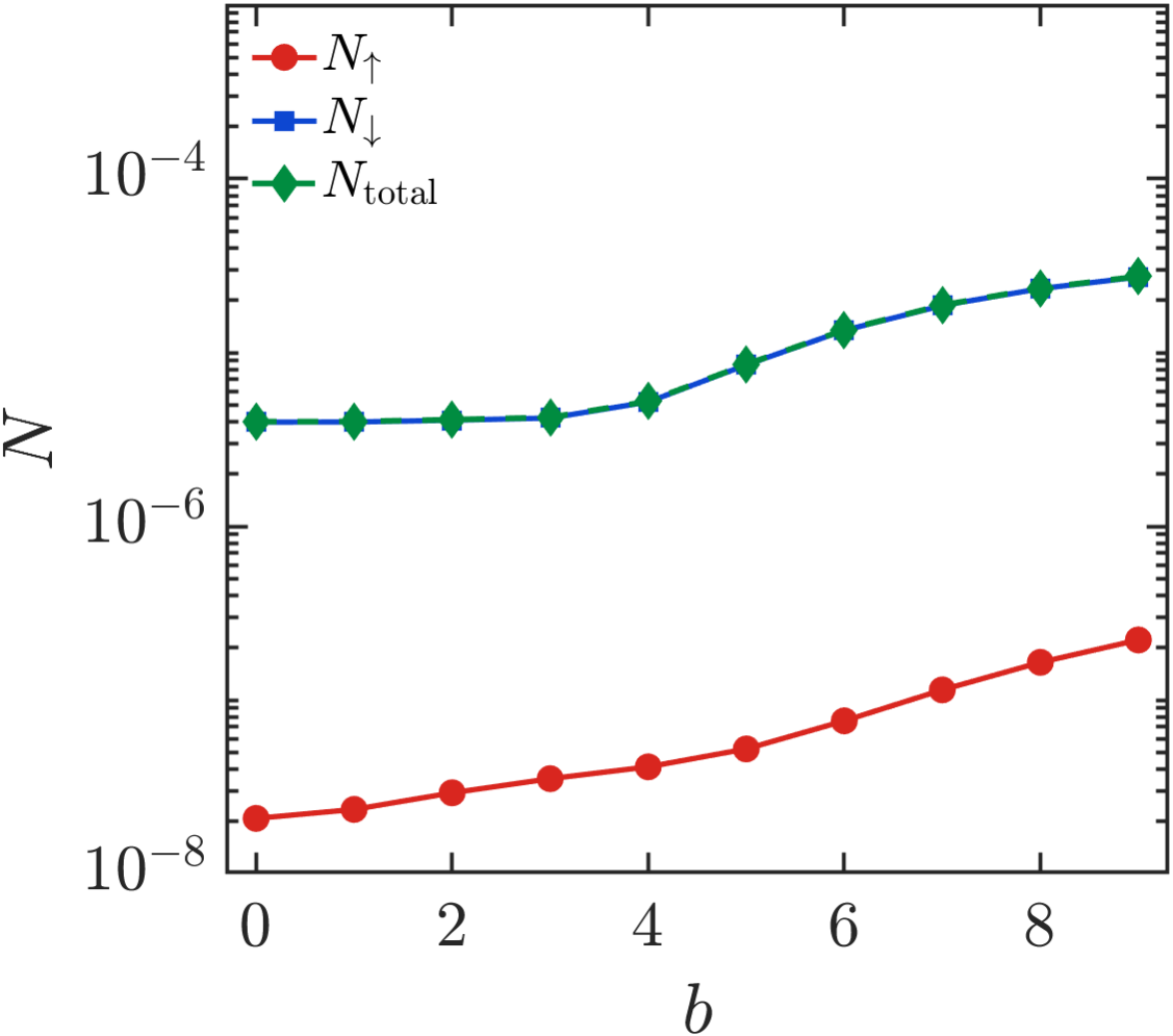}
			\captionsetup{justification=centering, singlelinecheck=true}
			\caption{Spin-resolved yields.}
			\label{fig:yields_amplitude}
		\end{subfigure}
		\hfill
		\begin{subfigure}[b]{0.48\linewidth}
			\centering
			\includegraphics[width=\linewidth]{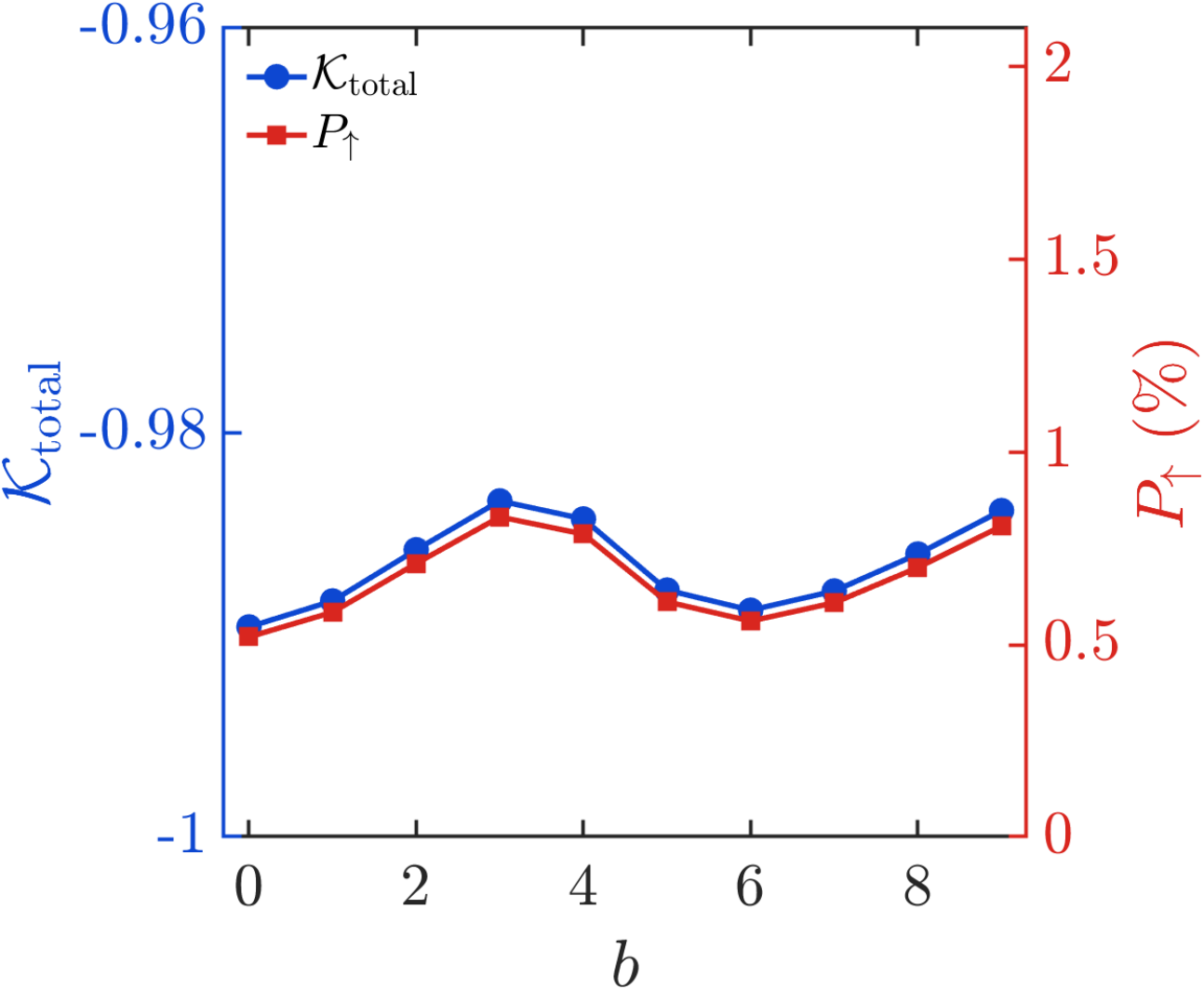}
			\captionsetup{justification=centering, singlelinecheck=true}
			\caption{Spin asymmetry and spin-up fraction.}
			\label{fig:asymmetry_amplitude}
		\end{subfigure}
		
		\caption{Spin-resolved yields and spin asymmetry as functions of the modulation amplitude \(b\) in a circularly polarized field with cosine-type phase modulation. The modulation frequency is fixed at $\Omega=0.04$; the other parameters are the same as in Fig.~\ref{fig:1}.}
		\label{fig:yields_asymmetry_amplitude}
	\end{figure}

	The dependence on \(b\) is shown in Fig.~\ref{fig:yields_asymmetry_amplitude}. The spin-resolved yields are shown in Fig.~\ref{fig:yields_amplitude}, and the corresponding spin asymmetry and spin-up fraction are shown in Fig.~\ref{fig:asymmetry_amplitude}. The spin-up and total yields increase with $b$ over the considered range, while the spin-down yield remains substantially larger than the spin-up yield. For \(\delta=+1\), the spin asymmetry varies only weakly and nonmonotonically with \(b\), while the spin-down channel remains favored for all \(b\). 

\begin{figure}[htbp]
		\centering
		\captionsetup{justification=raggedright, singlelinecheck=false}
		\setlength{\tabcolsep}{0pt}
		\begin{subfigure}[b]{0.44\linewidth}
			\centering
			\includegraphics[width=\textwidth]{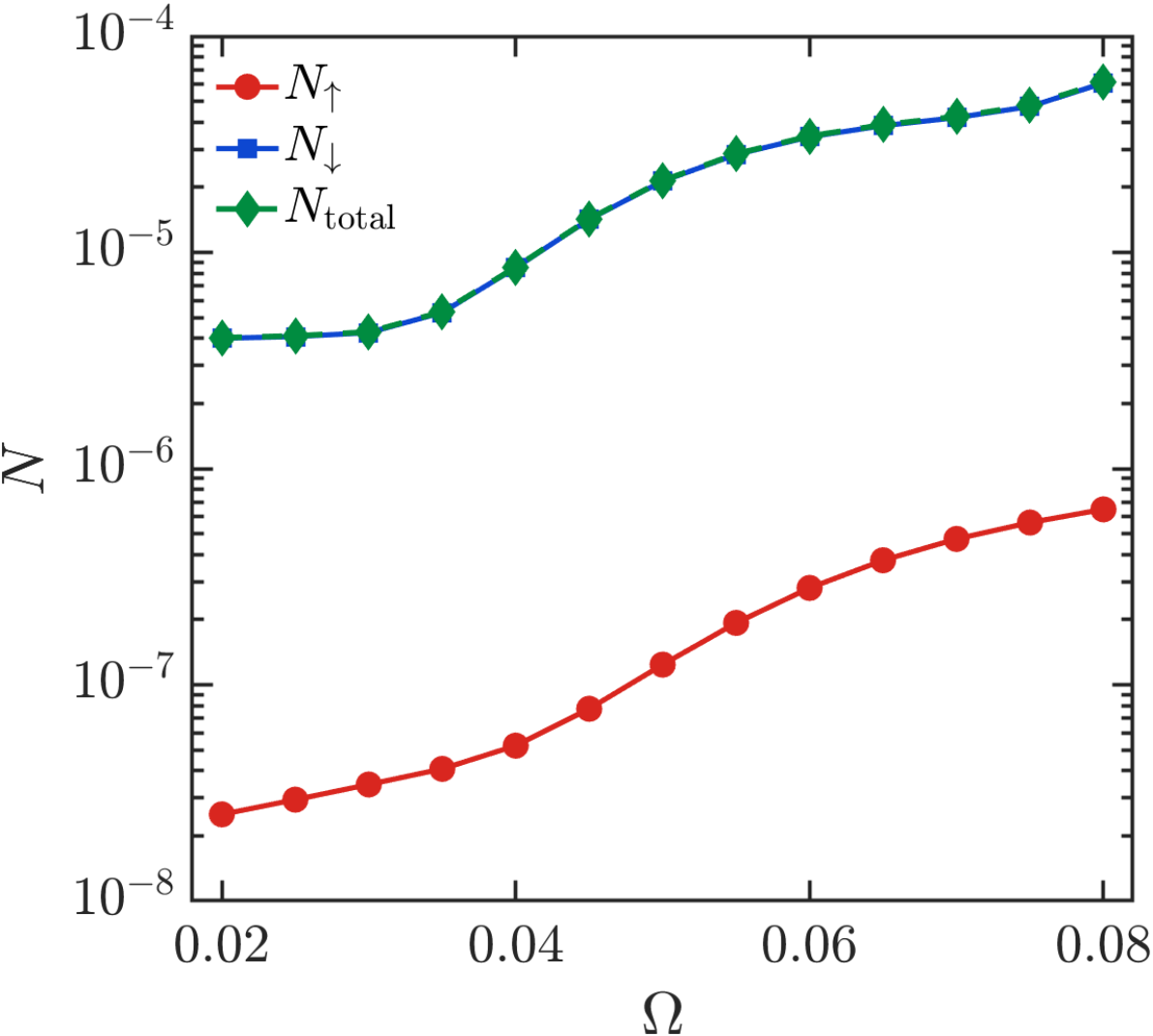}
			\captionsetup{justification=centering, singlelinecheck=true}
			\caption{Spin-resolved yields.}
			\label{fig:yields_frequency}
		\end{subfigure}
		\hfill
		\begin{subfigure}[b]{0.48\linewidth}
			\centering
			\includegraphics[width=\textwidth]{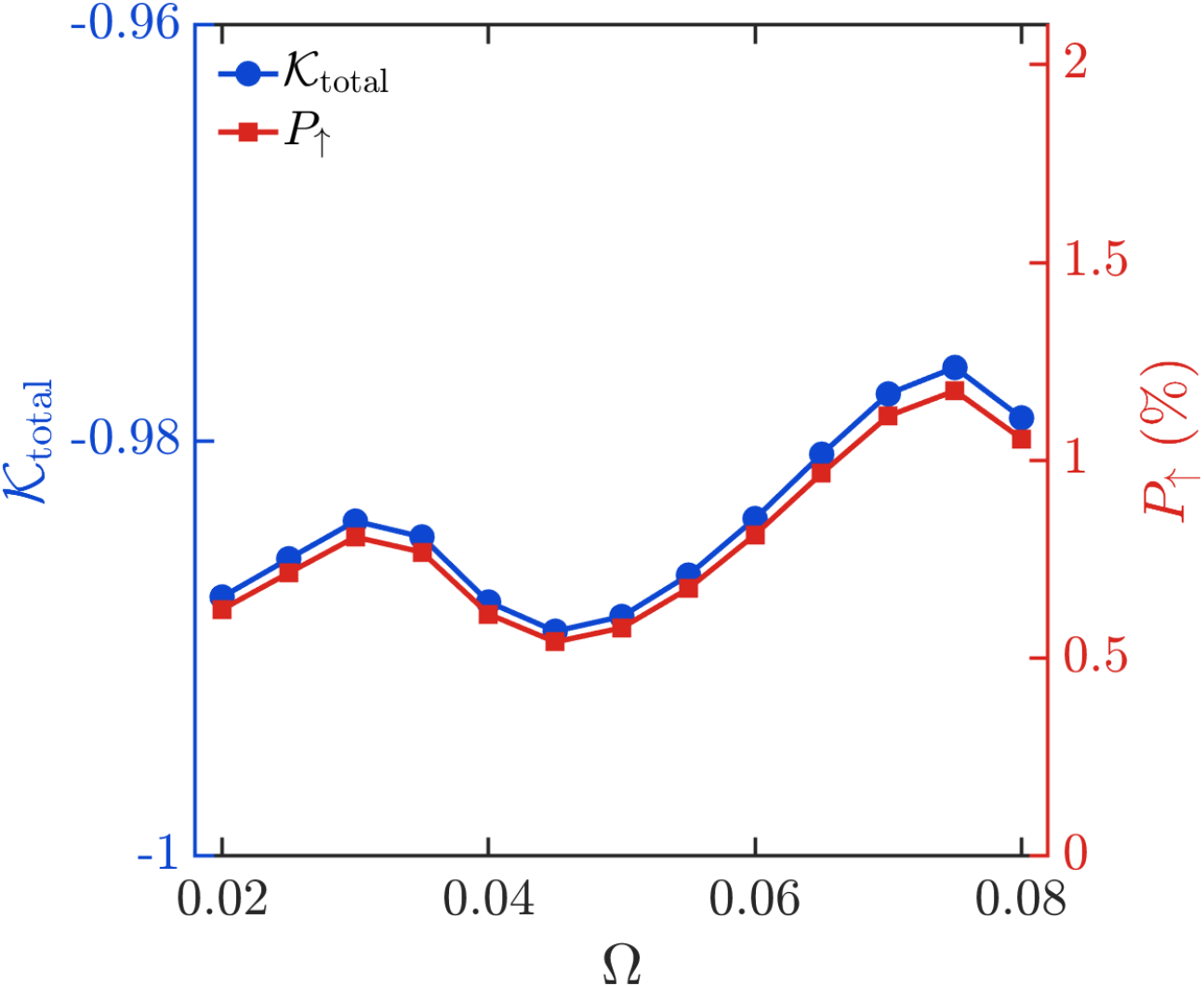}
			\captionsetup{justification=centering, singlelinecheck=true}
			\caption{Spin asymmetry and spin-up fraction.}
			\label{fig:asymmetry_frequency}
		\end{subfigure}
		
		\caption{Spin-resolved yields and spin asymmetry as functions of the modulation frequency \(\Omega\) in a circularly polarized field with cosine-type phase modulation. The modulation amplitude is fixed at $b=5$; the other parameters are the same as in Fig.~\ref{fig:1}.}\label{fig:yields_asymmetry_frequency}
	\end{figure}

	The dependence on \(\Omega\) is shown in
	Fig.~\ref{fig:yields_asymmetry_frequency}. The spin-resolved yields are shown in Fig.~\ref{fig:yields_frequency}, and the corresponding spin asymmetry and spin-up fraction are shown in Fig.~\ref{fig:asymmetry_frequency}. Both the spin-up and total yields grow with \(\Omega\) over the considered range. The spin-down yield provides the majority of the total yield, while the spin-up yield is considerably smaller. For \(\delta=+1\), the degree of the spin asymmetry varies weakly and nonmonotonically with \(\Omega\), while the spin-down channel remains favored for all \(\Omega\).
	
	For \(\delta=+1\), the modulation amplitude and frequency affect the degree of the spin asymmetry but not the favored spin channel, which remains spin-down. The asymmetry originates from the field rotation, the time-dependent phase modulation, and the coupled spin dynamics discussed below. The total yield and the spin asymmetry show different trends as \(b\) and \(\Omega\) vary, suggesting that pair production and spin imbalance are not simply proportional.
	
	To examine whether the favored spin channel is tied to the field handedness, we repeated the calculation with \(\delta=-1\). Table~\ref{tab:handedness} shows that at \(\Omega=0.04\) the favored channel reverses for both cosine-type and sine-type modulations, with \(N_{\uparrow}(\delta=-1)=N_{\downarrow}(\delta=+1)\) and \(N_{\downarrow}(\delta=-1)=N_{\uparrow}(\delta=+1)\) to numerical precision, while the total yield remains unchanged at fixed \(b\) and \(\Omega\). The same reversal is found for the cosine-type modulation at \(\Omega=0.02\), \(0.05\), \(0.06\), and \(0.08\). The magnitude \(|\mathcal{K}_{\mathrm{total}}|\) stays between \(0.96\) and \(0.99\) in all cases.
	
	\begin{table}[htbp]
		\centering
		\caption{Spin asymmetry \(\mathcal{K}_{\mathrm{total}}\) for opposite field handedness at \(\Omega=0.04\).}
		\label{tab:handedness}
		\footnotesize
		\setlength{\tabcolsep}{8pt}
		\begin{tabular}{cccc}
			\hline\hline
			Modulation & \(b\) & \(\mathcal{K}_{\mathrm{total}}(\delta=+1)\) & \(\mathcal{K}_{\mathrm{total}}(\delta=-1)\) \\
			\hline
			Cosine & 1 & \(-9.883\times10^{-1}\) & \(+9.883\times10^{-1}\) \\
			Cosine & 2 & \(-9.858\times10^{-1}\) & \(+9.858\times10^{-1}\) \\
			Cosine & 6 & \(-9.888\times10^{-1}\) & \(+9.888\times10^{-1}\) \\
			Cosine & 9 & \(-9.838\times10^{-1}\) & \(+9.838\times10^{-1}\) \\
			\hline
			Sine   & 1 & \(-9.856\times10^{-1}\) & \(+9.856\times10^{-1}\) \\
			Sine   & 2 & \(-9.759\times10^{-1}\) & \(+9.759\times10^{-1}\) \\
			Sine   & 6 & \(-9.879\times10^{-1}\) & \(+9.879\times10^{-1}\) \\
			Sine   & 9 & \(-9.652\times10^{-1}\) & \(+9.652\times10^{-1}\) \\
			\hline\hline
		\end{tabular}
	\end{table}
	
	We further qualitatively interpret the spin asymmetry using the effective magnetic field of a moving particle. In the QKEs, the spin-vector evolution is $\dot{\bm{F}}=2\bm{\mu}_1\times\bm{F}-2\bm{\mu}_2\times\bm{v}$, with $\bm{\mu}_1=\bm{p}\times\bm{E}/[2\varepsilon_{\bm{p}}(\varepsilon_{\bm{p}}+1)]$. Defining $\mathbf{B}_{\mathrm{eff}}\equiv 2\bm{\mu}_1$, the spin-precession equation becomes $\dot{\bm{F}}=\mathbf{B}_{\mathrm{eff}}\times\bm{F}-2\bm{\mu}_2\times\bm{v}$. The source term $-2\bm{\mu}_2\times\bm{v}$ excites $F_z$ from zero, while $\mathbf{B}_{\mathrm{eff}}$ drives the spin precession. Their interplay determines the momentum-space structure of $F_z(\bm q)$ and hence the spin asymmetry. 
	
	In the nonrelativistic limit, the effective field reduces to $\mathbf{B}_{\mathrm{eff}}\to\frac{1}{2}\bm{v}_{\mathrm{cl}}\times\bm{E}$, where $\bm{v}_{\mathrm{cl}}=\bm{p}/\varepsilon_{\bm{p}}$ is the classical velocity. This differs from the motional field \(-\bm{v}_{\mathrm{cl}}\times\bm{E}\) by a factor of \(-1/2\), which reflects the relativistic Thomas contribution in the precession rate of the QKEs. This result is consistent with the difference between the Foldy--Wouthuysen expansion and the leading-order Zeeman shift~\cite{wollert2015spin}. The effective field couples to the spin through $H_Z\simeq-\bm{\mu}\cdot\mathbf{B}_{\mathrm{eff}}$, where $\bm{\mu}$ is the spin magnetic moment. Because the spin magnetic moment $\bm{\mu}$ has opposite projections along $\mathbf{B}_{\mathrm{eff}}$ for spin-up and spin-down states, the two states experience opposite energy shifts. Since $\mathbf{B}_{\mathrm{eff}}$ depends on momentum, particles with different momenta accumulate different spin phases and thus different production probabilities. The favored spin channel is determined by the field handedness, which remains unchanged under variations of \(b\) and \(\Omega\) in the parameter range considered. This is consistent with the \(\delta=\pm1\) comparison in Table~\ref{tab:handedness}.

	\section{Effects of the phase-modulation form}\label{sec:phase_modulation_form}
\begin{figure}[htbp]
		\centering
		\captionsetup{justification=raggedright, singlelinecheck=false}
		\setlength{\tabcolsep}{0pt}
		\makebox[\textwidth][c]{%
			\makebox[0.05\textwidth][r]{}%
			\hspace{0.01\textwidth}%
			\begin{minipage}{0.27\textwidth}
				\centering
				{\small Spin-summed}
			\end{minipage}
			\hspace{0.01\textwidth}
			\begin{minipage}{0.27\textwidth}
				\centering
				{\small Spin-up}
			\end{minipage}
			\hspace{0.01\textwidth}
			\begin{minipage}{0.27\textwidth}
				\centering
				{\small Spin-down}
			\end{minipage}%
		}
		
		\vspace{0.2em}
		
		\makebox[\textwidth][c]{%
			\makebox[0.05\textwidth][r]{\small{\(b=1\)}}%
			\hspace{0.01\textwidth}%
			\begin{minipage}{0.27\textwidth}
				\centering
				\includegraphics[width=\linewidth]{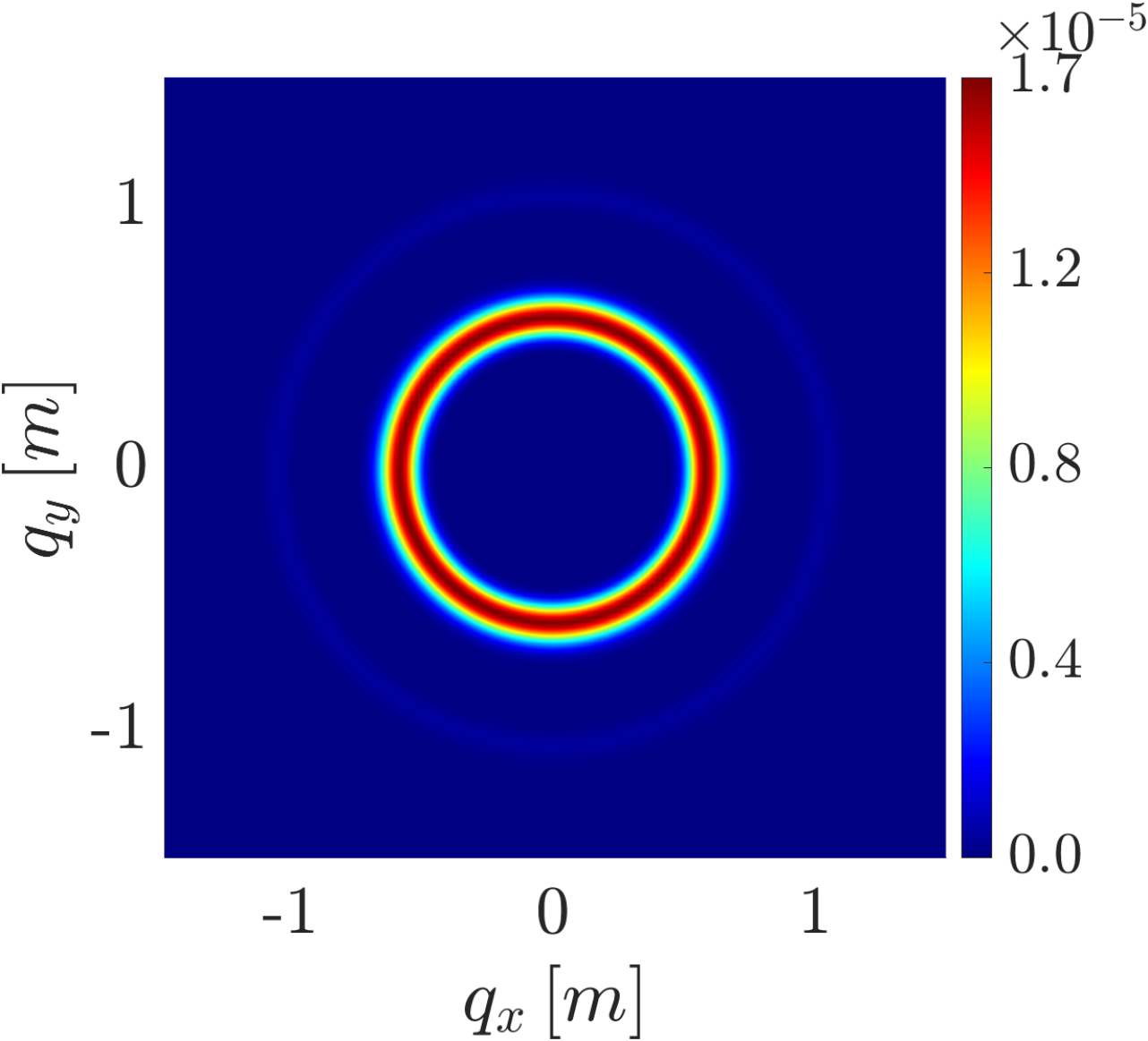}
			\end{minipage}
			\hspace{0.01\textwidth}
			\begin{minipage}{0.27\textwidth}
				\centering
				\includegraphics[width=\linewidth]{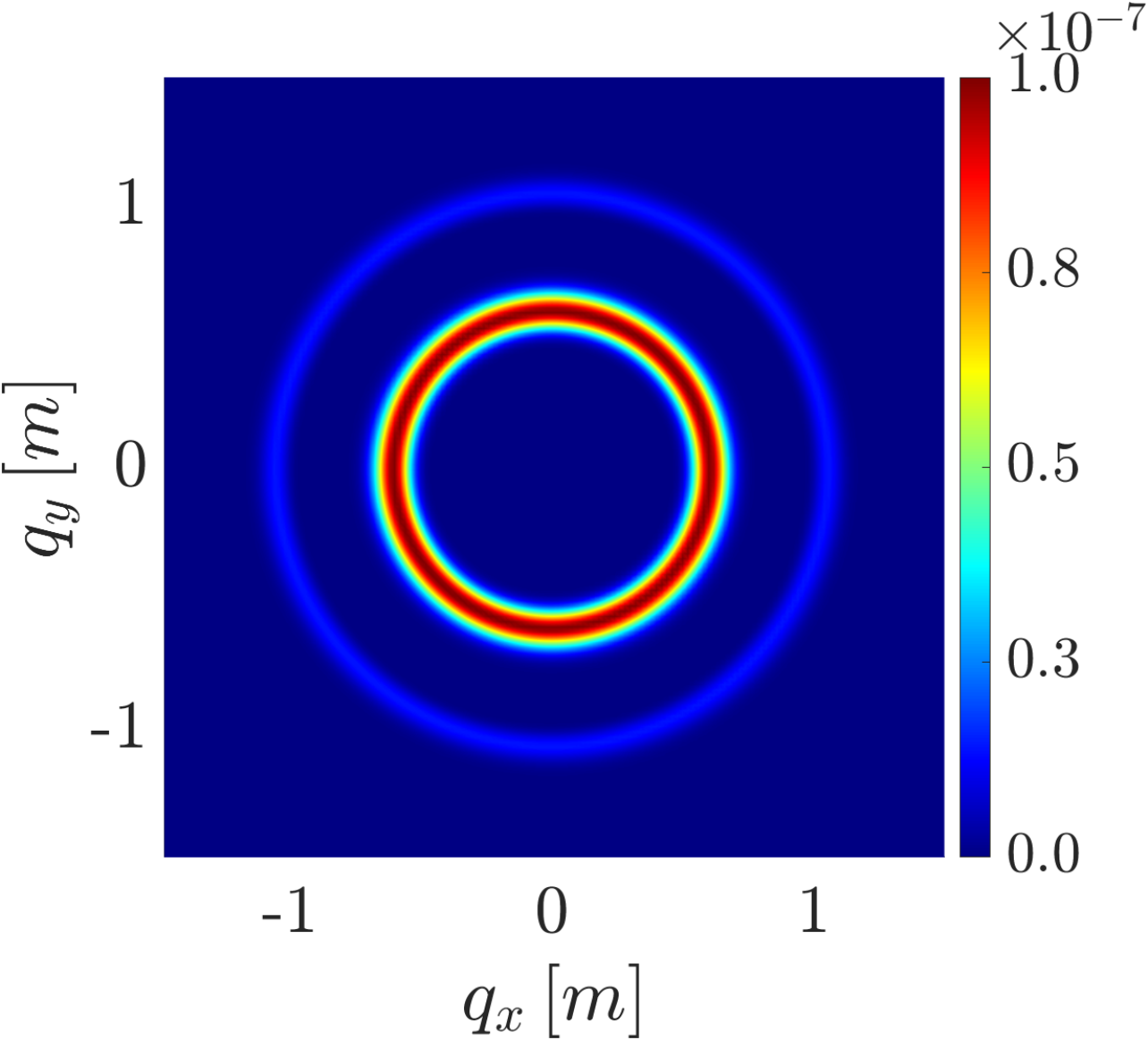}
			\end{minipage}
			\hspace{0.01\textwidth}
			\begin{minipage}{0.27\textwidth}
				\centering
				\includegraphics[width=\linewidth]{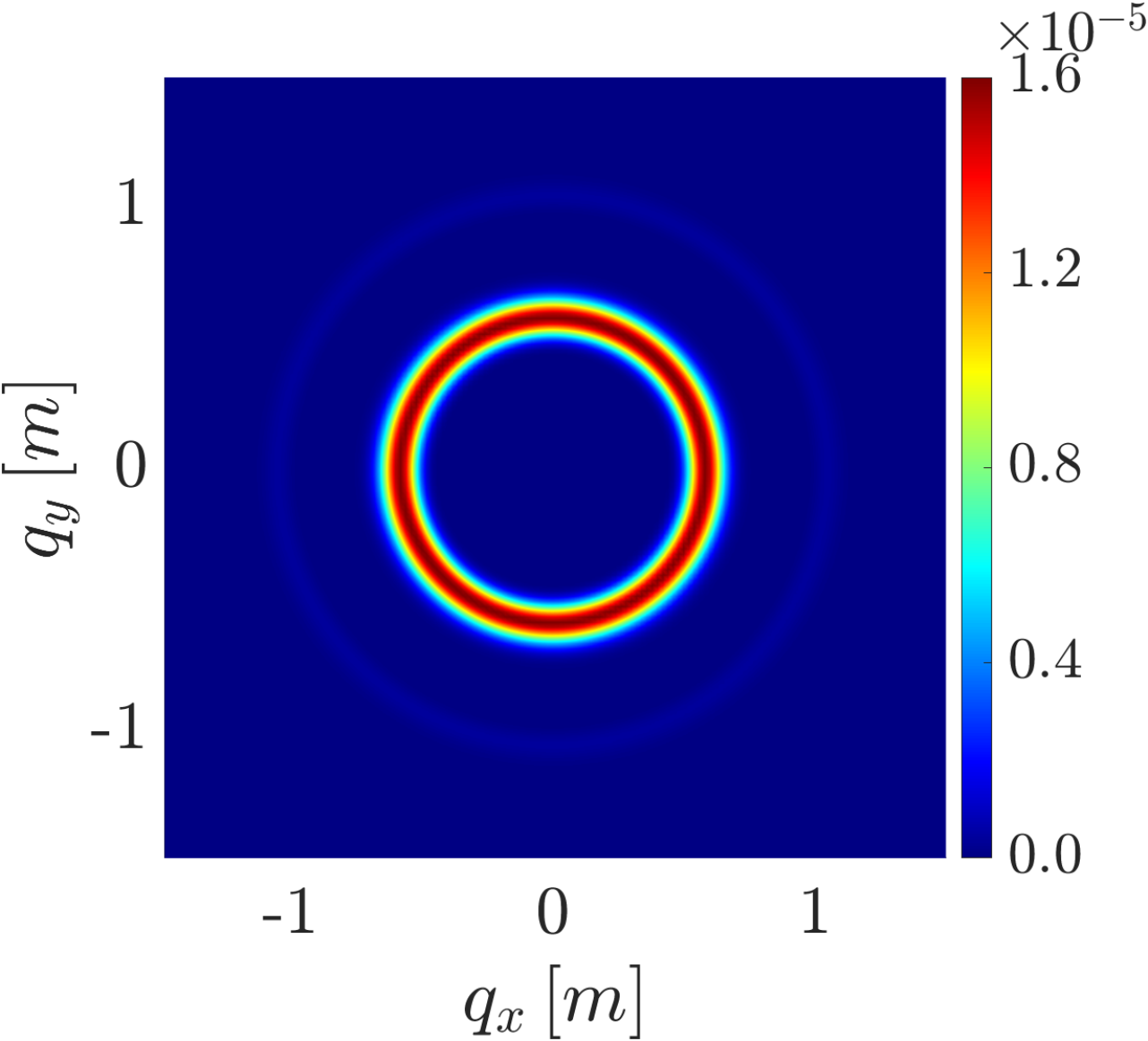}
			\end{minipage}%
		}
		
		\vspace{0.2em}
		
		\makebox[\textwidth][c]{%
			\makebox[0.05\textwidth][r]{\small{\(b=2\)}}%
			\hspace{0.01\textwidth}%
			\begin{minipage}{0.27\textwidth}
				\centering
				\includegraphics[width=\linewidth]{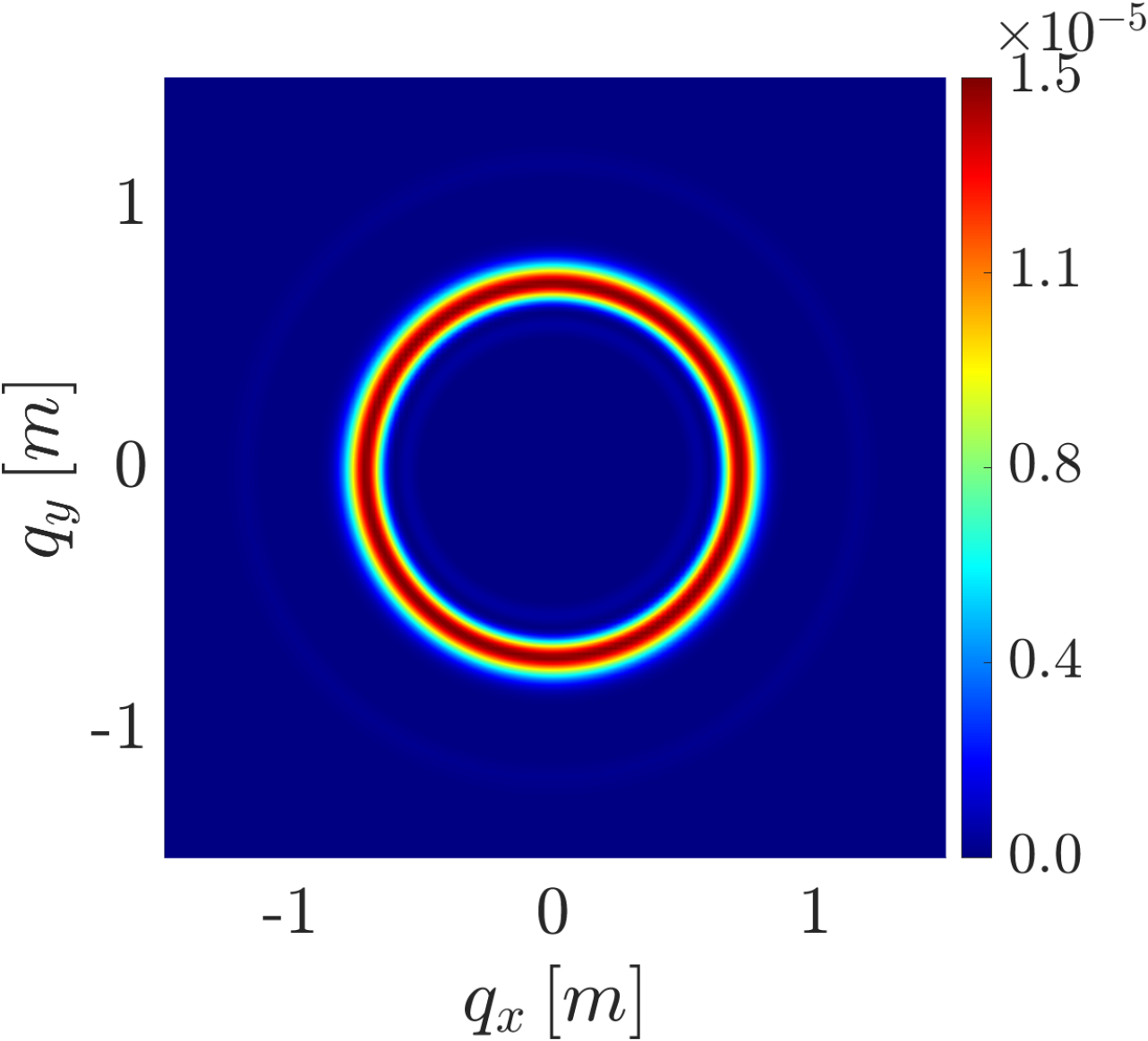}
			\end{minipage}
			\hspace{0.01\textwidth}
			\begin{minipage}{0.27\textwidth}
				\centering
				\includegraphics[width=\linewidth]{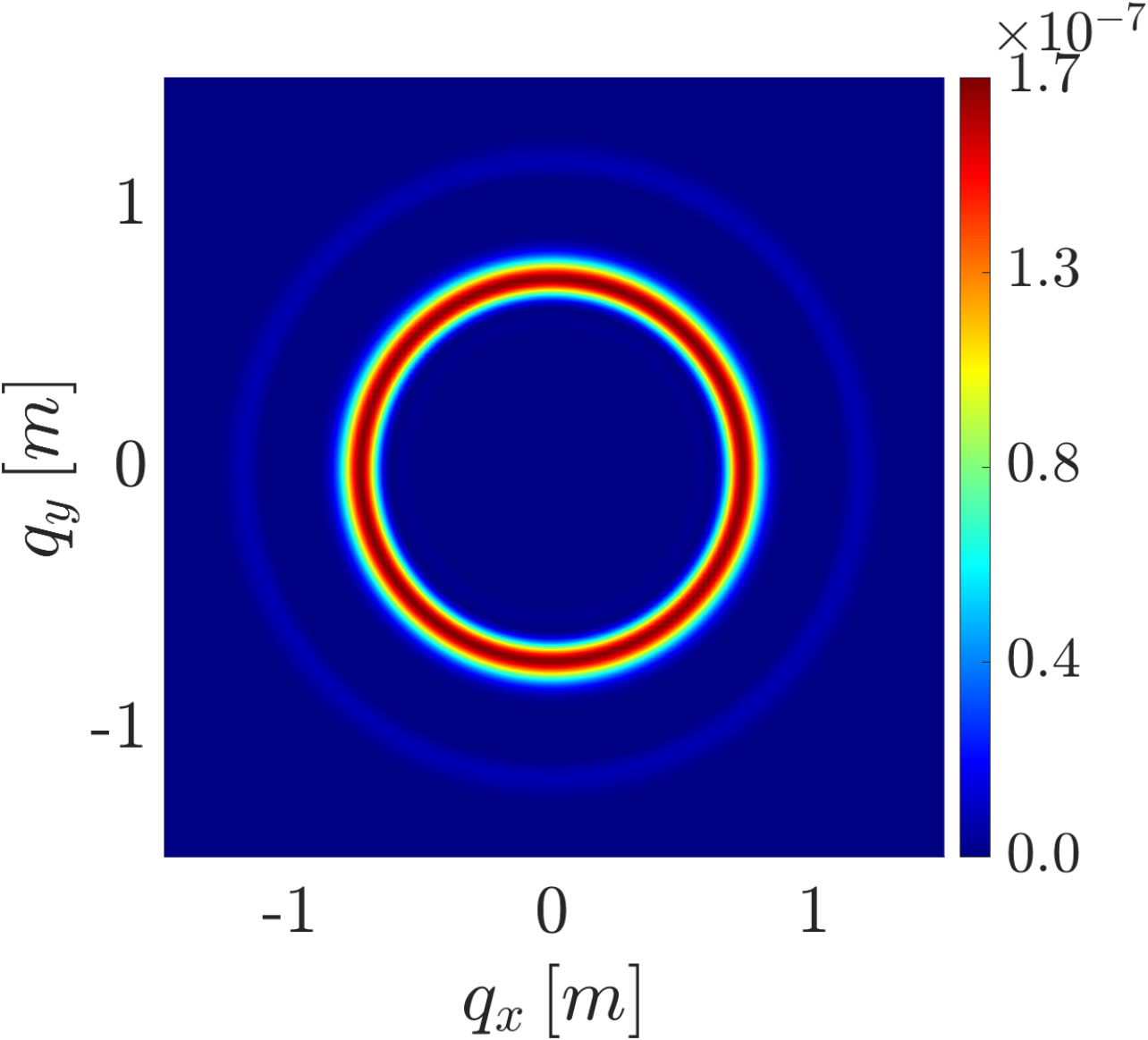}
			\end{minipage}
			\hspace{0.01\textwidth}
			\begin{minipage}{0.27\textwidth}
				\centering
				\includegraphics[width=\linewidth]{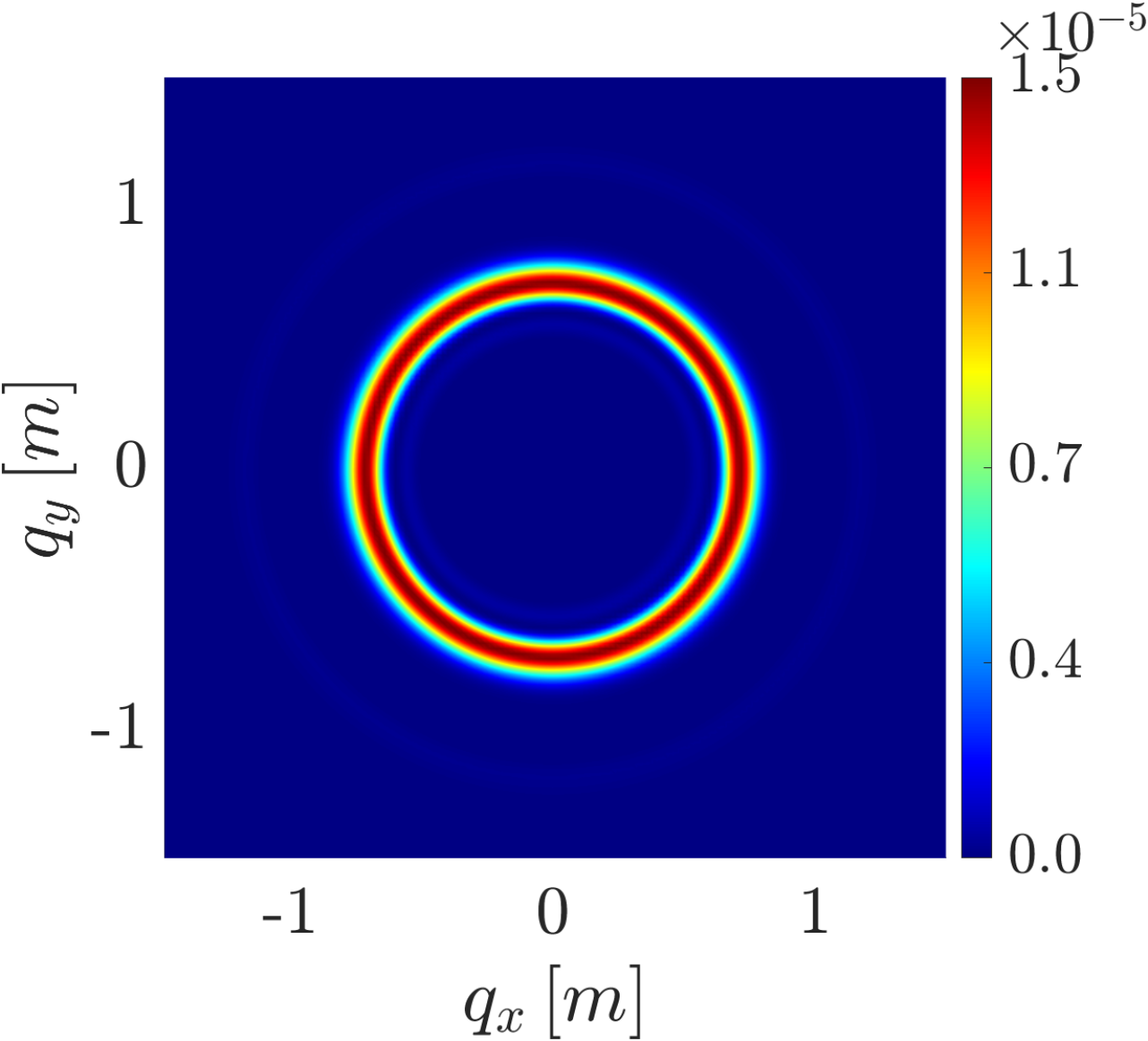}
			\end{minipage}%
		}
		
		\vspace{0.2em}
		
		\makebox[\textwidth][c]{%
			\makebox[0.05\textwidth][r]{\small{\(b=6\)}}%
			\hspace{0.01\textwidth}%
			\begin{minipage}{0.27\textwidth}
				\centering
				\includegraphics[width=\linewidth]{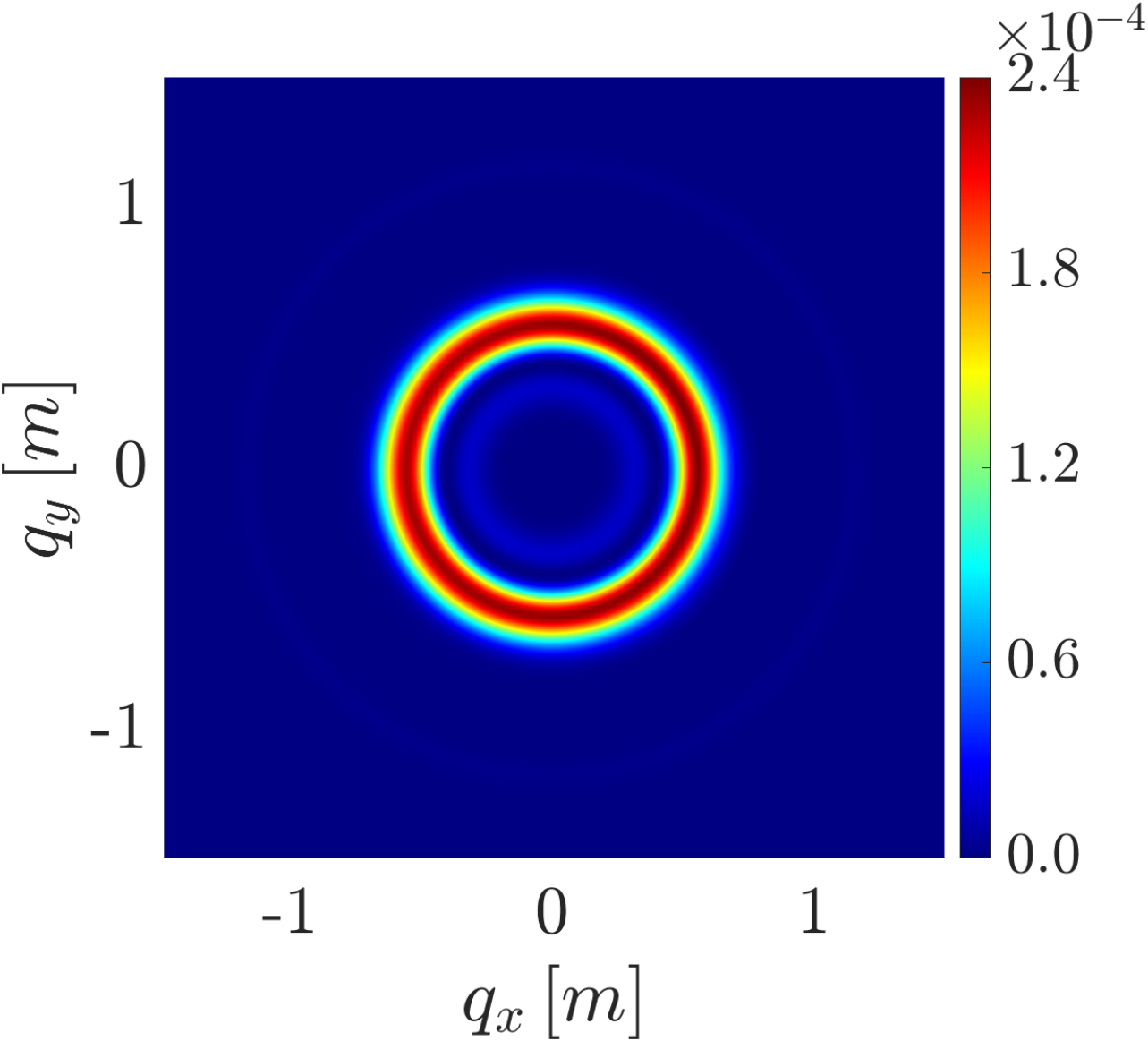}
			\end{minipage}
			\hspace{0.01\textwidth}
			\begin{minipage}{0.27\textwidth}
				\centering
				\includegraphics[width=\linewidth]{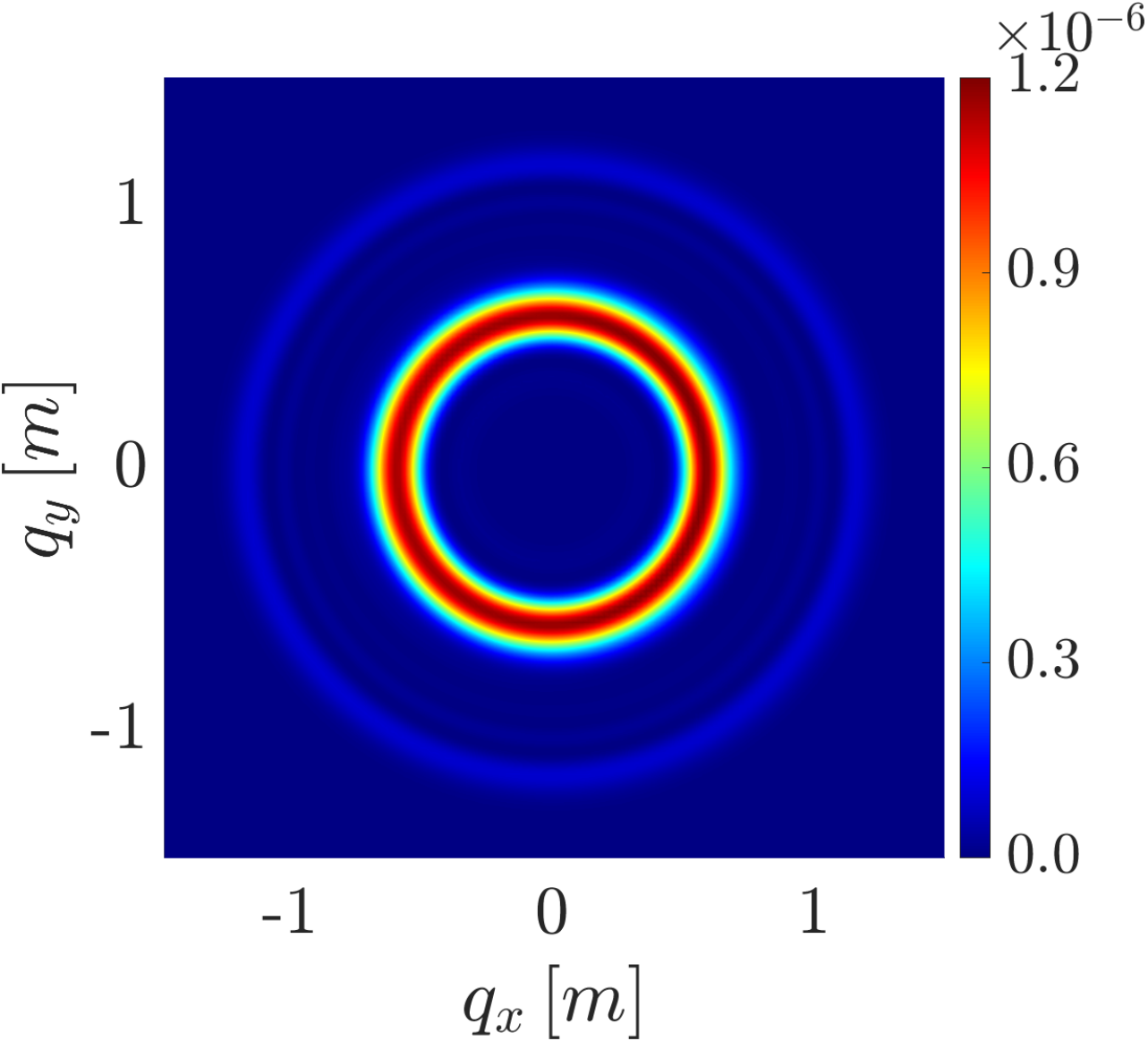}
			\end{minipage}
			\hspace{0.01\textwidth}
			\begin{minipage}{0.27\textwidth}
				\centering
				\includegraphics[width=\linewidth]{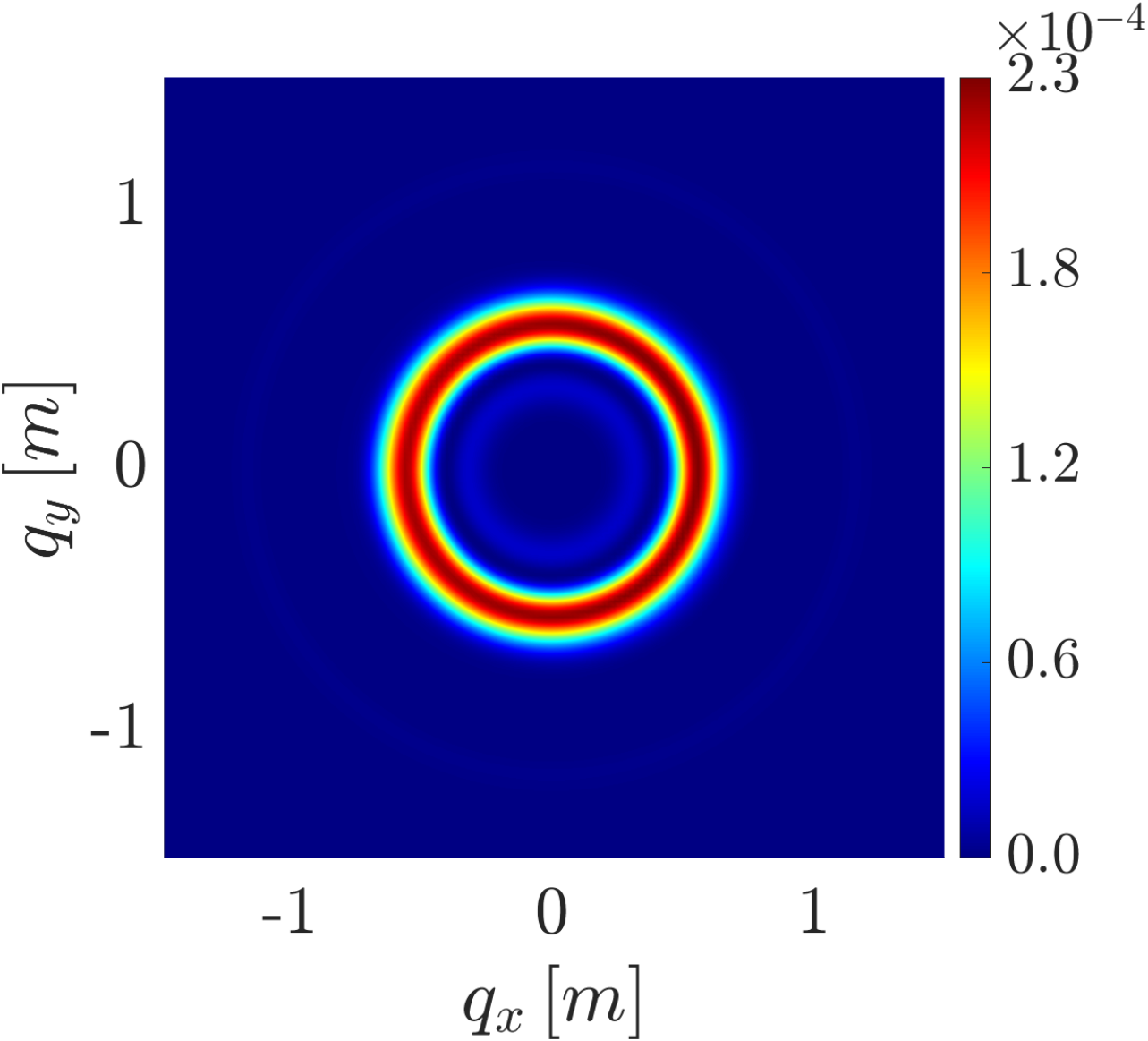}
			\end{minipage}%
		}
		
		\vspace{0.2em}
		
		\makebox[\textwidth][c]{%
			\makebox[0.05\textwidth][r]{\small{\(b=9\)}}%
			\hspace{0.01\textwidth}%
			\begin{minipage}{0.27\textwidth}
				\centering
				\includegraphics[width=\linewidth]{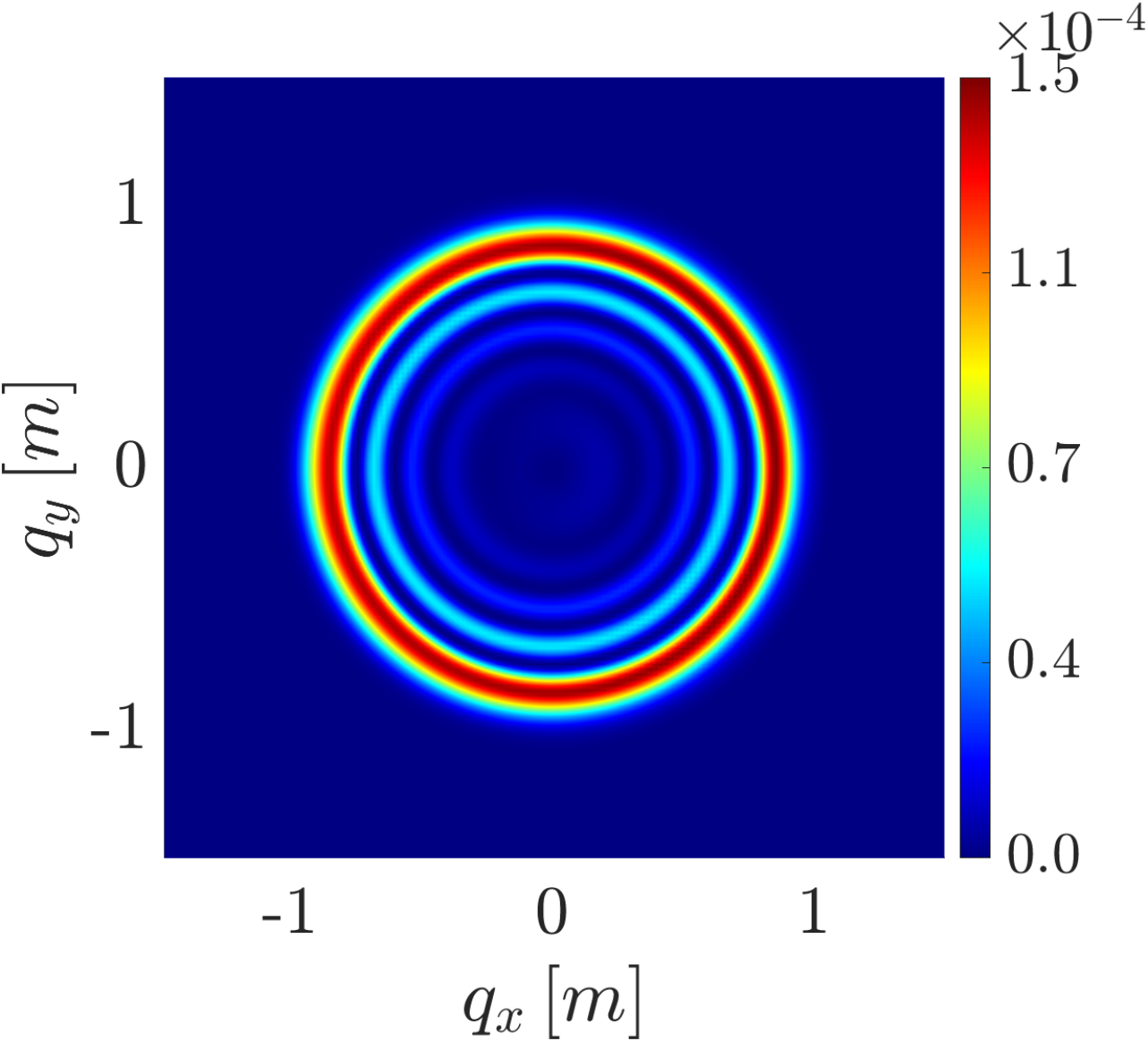}
			\end{minipage}
			\hspace{0.01\textwidth}
			\begin{minipage}{0.27\textwidth}
				\centering
				\includegraphics[width=\linewidth]{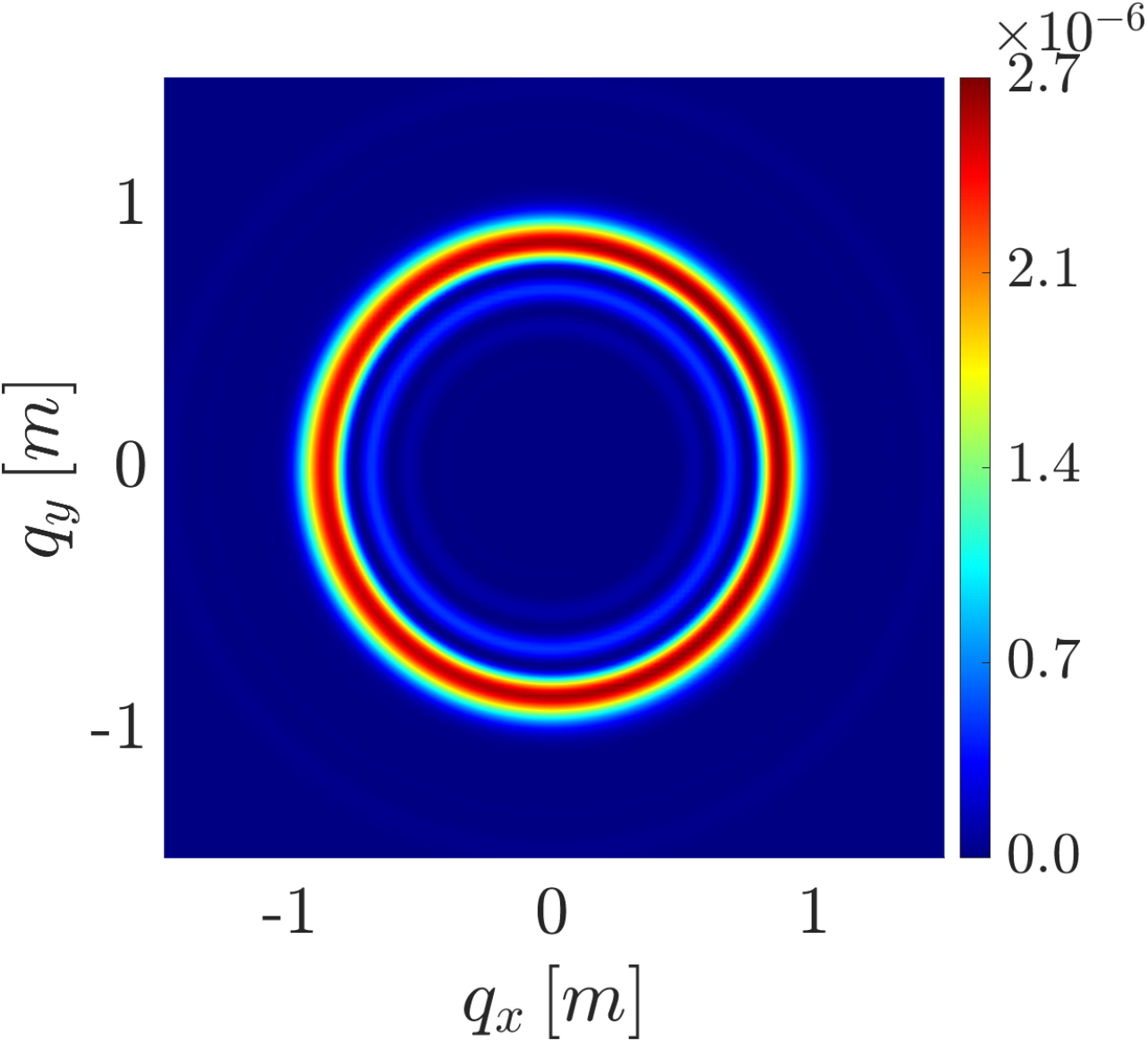}
			\end{minipage}
			\hspace{0.01\textwidth}
			\begin{minipage}{0.27\textwidth}
				\centering
				\includegraphics[width=\linewidth]{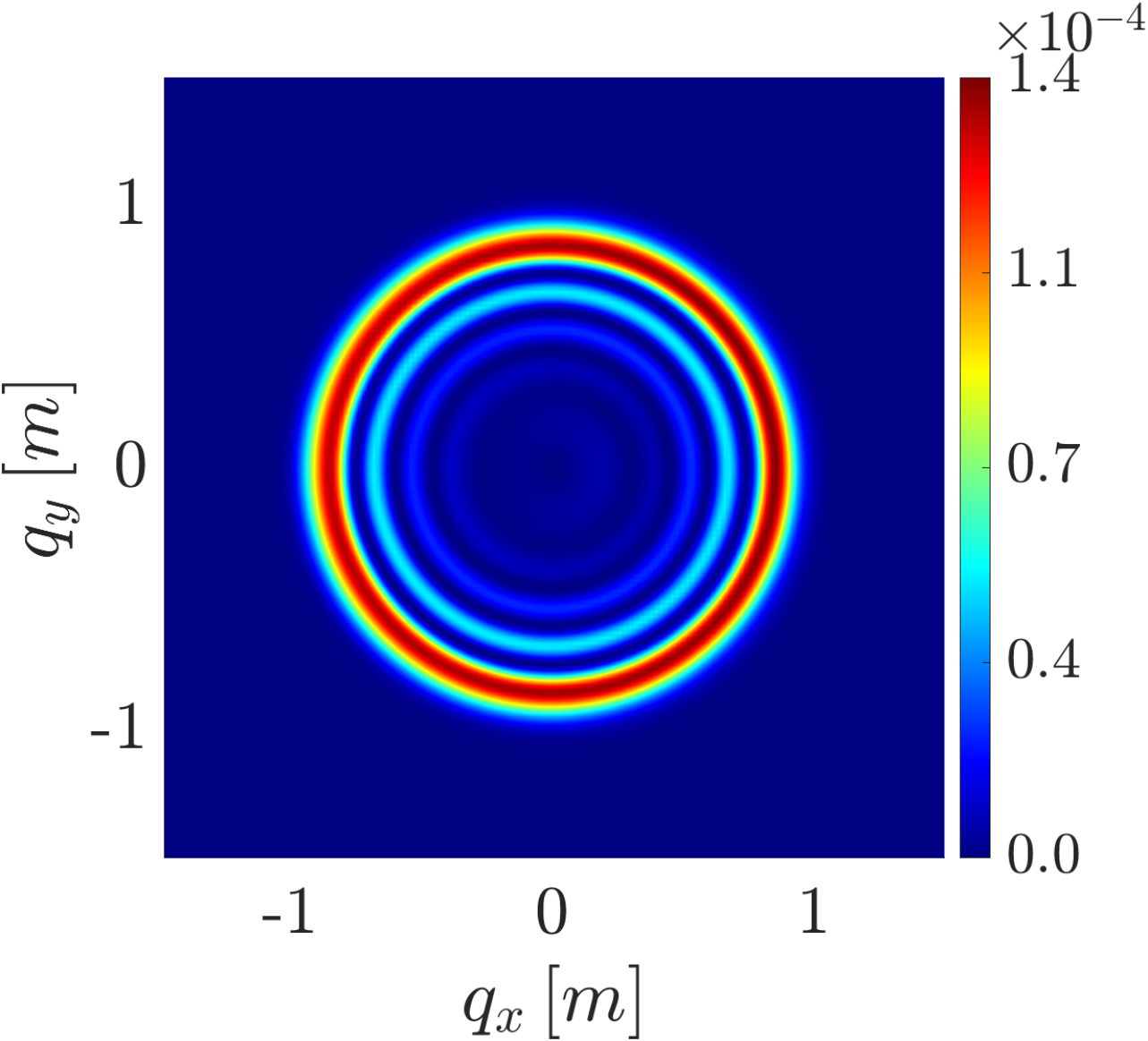}
			\end{minipage}%
		}
		
		\caption{Transverse momentum distributions of spin-summed, spin-up, and spin-down particles for the sine-type phase modulation in a circularly polarized field. The modulation frequency is \(\Omega=0.04\), the modulation amplitudes are \(b=1\), \(2\), \(6\), and \(9\), and \(q_z=0\). The remaining parameters are the same as in Fig.~\ref{fig:1}.}
		\label{fig:sin_spectra}
	\end{figure}

	We now study how the form of the phase modulation affects the momentum distributions and spin asymmetry, comparing the cosine-type and sine-type modulations. The remaining parameters are the same as in Sec.~\ref{sec:results}. The sine-type results for the spin-summed, spin-up, and spin-down
	distributions are shown in Fig.~\ref{fig:sin_spectra}. The corresponding cosine-type spin-up distributions were discussed in Sec.~\ref{subsec:modulation_amplitude}.
	
	For the sine-type modulation, the three distributions remain predominantly ring-shaped and approximately mirror-symmetric about the \(q_y\) axis as \(b\) increases. The two spin-resolved
	distributions have similar momentum-space patterns but different
	magnitudes. In contrast, the cosine-type spin-up distributions show stronger directional deformation and a more pronounced spiral structure, particularly at larger \(b\). This difference originates from the temporal symmetry of the field. For the sine-type modulation, \(E_x(t)\) is even and \(E_y(t)\) is odd under \(t\to -t\). With the integration constant chosen such that \(A_x(0)=A_y(0)=0\), the vector potential satisfies \(A_x(-t)=-A_x(t)\) and \(A_y(-t)=A_y(t)\). Under the transformation \(t\to-t\) together with \(q_x\to-q_x\), the term \(q_x-eA_x(t)\) changes sign while \(q_y-eA_y(t)\) is unchanged, so the instantaneous energy
	\(\varepsilon(\bm q,t)=\sqrt{1+[q_x-eA_x(t)]^2+[q_y-eA_y(t)]^2}\) remains invariant. This is consistent with the approximately mirror-symmetric distribution about the \(q_y\) axis. For the cosine-type modulation, these field components are neither even nor odd, so this symmetry is broken and the distribution can develop stronger directional deformation and spiral structure.

\begin{figure}[htbp]
		\centering
		\captionsetup{justification=raggedright, singlelinecheck=false}
		\setlength{\tabcolsep}{0pt}
		
		\begin{subfigure}[b]{0.47\linewidth}
			\centering
			\includegraphics[width=\linewidth]{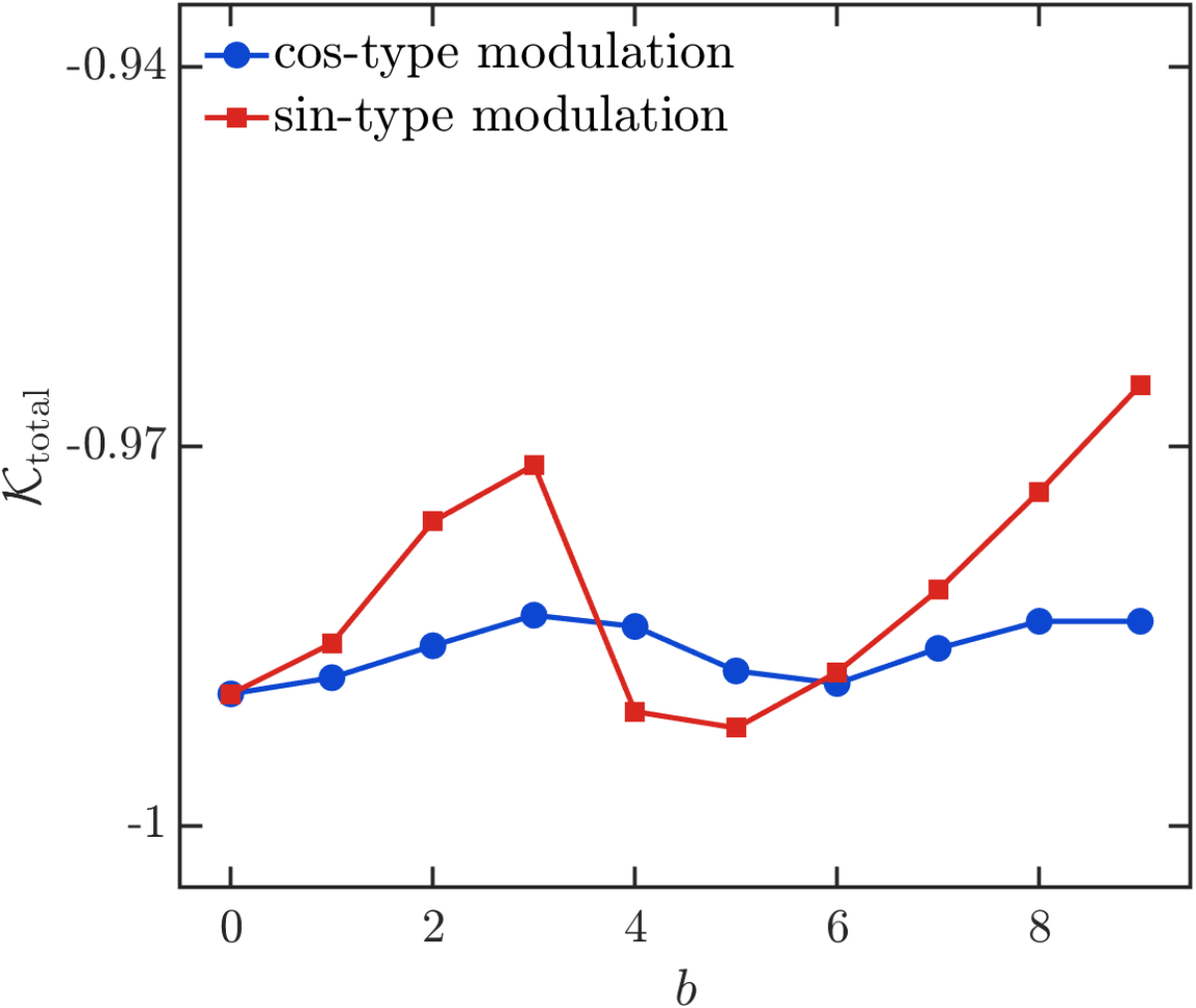}
			\captionsetup{justification=centering, singlelinecheck=true}
			\caption{Spin asymmetry \(\mathcal{K}_{\mathrm{total}}\).}
		\end{subfigure}
		\hfill
		\begin{subfigure}[b]{0.45\linewidth}
			\centering
			\includegraphics[width=\linewidth]{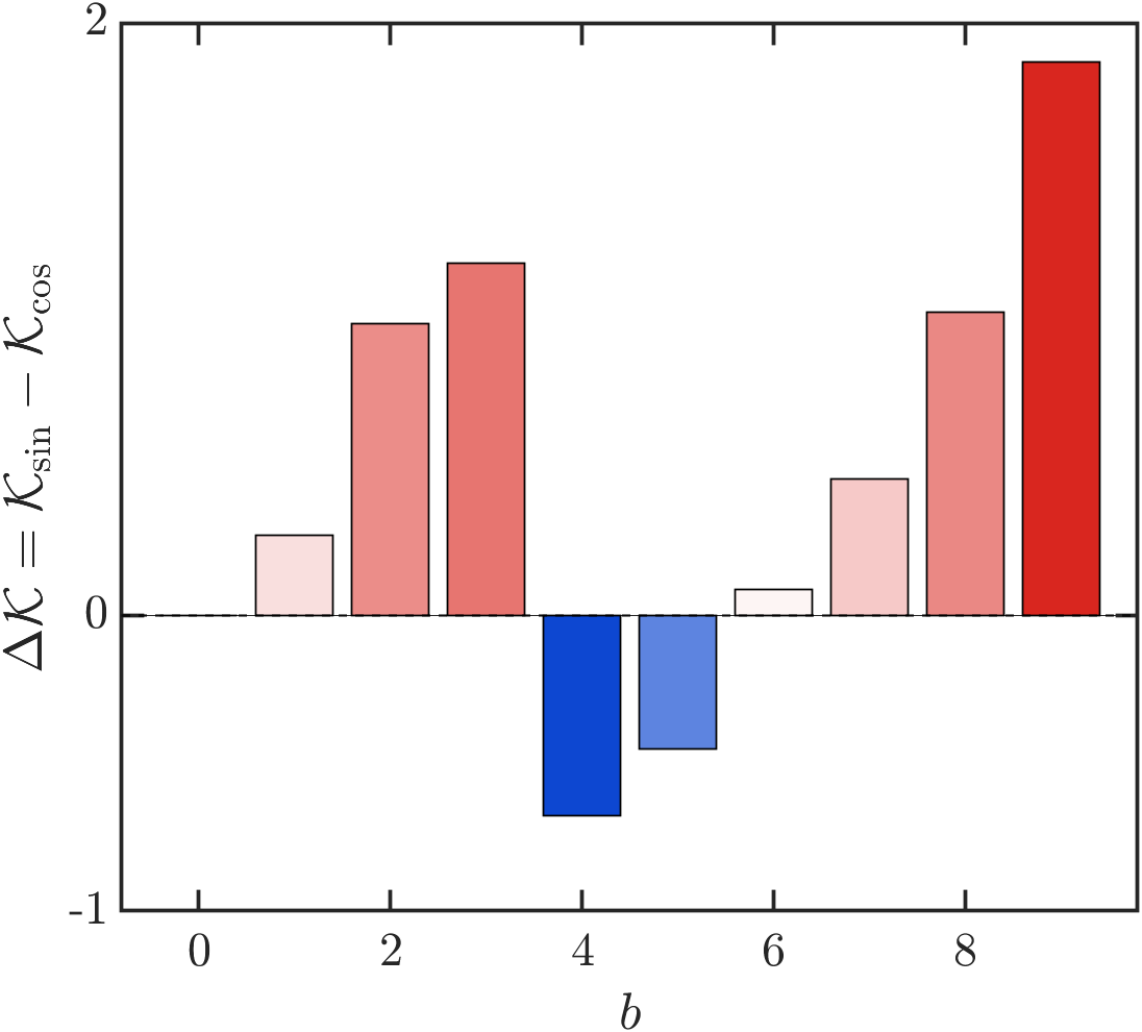}
			\captionsetup{justification=centering, singlelinecheck=true}
			\caption{Difference \(\Delta\mathcal{K}_{\mathrm{total}}\).}
		\end{subfigure}
		
		\caption{Spin asymmetry and its difference for cosine-type and sine-type phase modulations in a circularly polarized field with \(\Omega=0.04\). 
			(a) The total spin asymmetry \(\mathcal{K}_{\mathrm{total}}\) as a function of the modulation amplitude \(b\). 
			(b) The difference \(\Delta\mathcal{K}_{\mathrm{total}} = \mathcal{K}_{\mathrm{total}}^{\mathrm{sin}} - \mathcal{K}_{\mathrm{total}}^{\mathrm{cos}}\) versus \(b\), where red and blue bars correspond to positive and negative values, respectively, and the shading intensity reflects their magnitude.}
		\label{fig:asymmetry_comparison}
	\end{figure}

	We now compare the spin asymmetry produced by the two modulation forms. The dependence of \(\mathcal{K}_{\mathrm{total}}\) on \(b\) is shown in Fig.~\ref{fig:asymmetry_comparison}(a). As shown in Sec.~\ref{subsec:spin_resolved_yields}, the spin asymmetry for the cosine-type modulation varies only weakly and nonmonotonically with \(b\). In contrast, the sine-type modulation shows a stronger dependence on \(b\): the spin-down dominance is weaker at \(b=1\) and \(b=2\), remains strong at \(b=6\), and decreases again at \(b=9\). The cosine-type modulation strongly affects the directional structure while leaving the asymmetry stable, whereas the sine-type modulation preserves the mirror-symmetric pattern but makes the asymmetry strongly dependent on \(b\). The phase-modulation form therefore affects the spin-resolved observables through both the momentum-space geometry and the relative weighting of the two spin channels. Reversing the field handedness reverses the favored channel for the sine-type modulation as well, with the total yield unchanged (Table~\ref{tab:handedness}). The handedness--spin correspondence is therefore a property of the circularly polarized field itself and does not depend on the modulation form.
	
	The difference between the two forms is quantified by
	\begin{equation}
		\Delta\mathcal{K}_{\mathrm{total}}
		=
		\mathcal{K}_{\mathrm{total}}^{\mathrm{sin}}
		-
		\mathcal{K}_{\mathrm{total}}^{\mathrm{cos}},
	\end{equation}
	shown in Fig.~\ref{fig:asymmetry_comparison}(b). The difference changes sign as \(b\) varies: it is positive for \(b=1,2,3\), negative near \(b=4\) and \(b=5\), and positive again at \(b=6\) and \(b=9\). Neither form consistently produces a larger \(|\mathcal{K}_{\mathrm{total}}|\) over the entire parameter range, and their relative magnitudes depend sensitively on \(b\).
	
	The nonmonotonic behavior of \(\Delta\mathcal{K}_{\mathrm{total}}\) reflects the different temporal phase structures of the two modulations. Both fields have the same carrier frequency, pulse envelope, field strength, and ellipticity. However, the sine- and cosine-type modulations place the strong-field region at different phases of the field rotation. This modifies the coupling among the field, the particle motion, and the spin dynamics. The spin asymmetry is therefore sensitive not only to the overall field strength but also to the temporal phase structure of the pulse. The difference originates from the different time evolutions of the effective magnetic field \(\mathbf{B}_{\mathrm{eff}}=2\bm{\mu}_1\) experienced by a moving particle. The sine- and cosine-type modulations produce different instantaneous frequencies $\omega_{\mathrm{inst}}(t)$, and hence different field histories $\bm E(t)$ and different \(\mathbf{B}_{\mathrm{eff}}(t)\) throughout the pulse.

	\section{Summary and Outlook}
	\label{sec:summary}
	
	We have studied vacuum electron--positron pair production in phase-modulated elliptically polarized electric fields using the QKEs, examining how the modulation amplitude, frequency, ellipticity, and form of phase modulation affect the spin-resolved momentum spectra and the spin asymmetry. For circular polarization with cosine-type modulation, increasing the modulation amplitude deforms ring-shaped distributions into spiral patterns. The deformation does not increase monotonically with either the modulation amplitude or the frequency, and is most pronounced at intermediate values within the sampled range. At larger amplitudes, the contour develops segments with different local orientations, and the local linear fits become less reliable. As the ellipticity increases, the distribution becomes less elongated and more ring-like, and the spiral deformation becomes more pronounced; at \(\delta=1\), the transverse motion is most isotropic, so the modulation-induced phase differences directly reshape the azimuthal distribution.
	
	The form of the phase modulation affects the spin asymmetry independently of the momentum-space geometry. For circular polarization with \(\delta=+1\), the spin-down channel is favored over the ranges of modulation amplitude and frequency considered in the \(q_z=0\) plane. Reversing the field handedness reverses the favored channel and leaves the total yield unchanged, for both cosine-type and sine-type modulations. Sine-type modulation keeps the distribution ring-shaped and approximately mirror-symmetric, but makes the asymmetry strongly dependent on the modulation amplitude. Since both modulations share the same carrier frequency, pulse envelope, field strength, and ellipticity, the difference must originate from their temporal phase structure, which places the strong-field region at different phases of the field rotation.
	
	Ellipticity thus sets the overall shape of the momentum distribution, while phase modulation controls the spiral structure and the degree of spin asymmetry. Because geometric deformation and spin imbalance are governed by different aspects of the field, they can be analyzed separately in spin-resolved studies. The present analysis does not explain the spiral structures dynamically. A complex-time semiclassical analysis could identify the dominant turning points and clarify the interference mechanisms behind the observed momentum patterns~\cite{dumlu2011complex}. 
	
	Two natural extensions are to consider the involved problem with phase modulation in spatially inhomogeneous fields and magnetic fields as well. Spatially inhomogeneous fields introduce additional geometric effects through the finite extent of realistic laser pulses, while magnetic fields couple directly to the spin and may change both the momentum spectra and the spin asymmetry. These problems, which are beyond present work, would be interesting and worthy to be studied in future furthermore. 
	
	\acknowledgments
	This work was supported by the National Natural Science Foundation of China under Grant Nos. 12375240 and 11935008. The computation was carried out at the
	HSCC of Beijing Normal University.
	
	\bibliography{references}
	
\end{document}